\documentclass[12pt,preprint]{aastex}

\newcommand{\iso}[2]{\hbox{${}^{#1}{\rm #2}$}}
\newcommand{\Msun}{\ensuremath{{M}_{\sun}}}

\usepackage{graphicx}

\shorttitle{Evolution \& Nucleosynthesis of AGB stars}
\shortauthors{D. Kamath, A.I. Karakas \& P.R. Wood}

\begin{document}


\title{Evolution and  Nucleosynthesis of AGB stars in Three Magellanic 
Cloud Clusters}


\author{D. Kamath}
\affil{Research School of Astronomy \& Astrophysics, Mount Stromlo Observatory,
Weston Creek ACT 2611, Australia}
\email{devika13@mso.anu.edu.au}

\author{A.I. Karakas\altaffilmark{1}}
\affil{Research School of Astronomy \& Astrophysics, Mount Stromlo Observatory,
Weston Creek ACT 2611, Australia}
\email{akarakas@mso.anu.edu.au}

\and

\author{P.R. Wood  }
\affil{Research School of Astronomy \& Astrophysics, Mount Stromlo Observatory,
Weston Creek ACT 2611, Australia}
\email{wood@mso.anu.edu.au}


\altaffiltext{1} {Centre for Stellar \& Planetary Astrophysics, Monash University,
Clayton VIC 3800, Australia}


\begin{abstract}

We present stellar evolutionary sequences for asymptotic giant branch
(AGB) stars in the Magellanic Cloud clusters NGC\,1978, NGC\,1846 
and NGC\,419. The new stellar models for the 
three clusters match the observed effective temperatures on 
the giant branches, the oxygen-rich to carbon-rich transition 
luminosities, and the AGB-tip luminosities. A major finding is that a large amount of 
convective overshoot (up to 3 pressure scale heights) is required at the base of the 
convective envelope during third dredge-up in order to get the correct 
oxygen-rich to carbon-rich transition luminosity. The stellar evolution sequences 
are used as input for detailed nucleosynthesis calculations. 
For NGC\,1978 and NGC\,1846 we compare our model results to the 
observationally derived abundances of carbon and oxygen. We find that
additional mixing processes (extra-mixing) are required to 
explain the observed abundance 
patterns. For NGC\,1846 we conclude that non-convective extra-mixing processes 
are required on both the RGB and the AGB, in agreement with previous studies. 
For NGC\,1978 it is possible to explain the C/O and \iso{12}C/\iso{13}C 
abundances of both the O-rich and the C-rich AGB stars by assuming that 
the material in the intershell region contains high abundances of both C and O. 
This may occur during a thermal pulse when 
convective overshoot at the inner edge of the flash-driven convective pocket 
dredges C and O from the core to the intershell. 
For NGC\,419 we provide our predicted model abundance 
values although there are currently no published observed abundance 
studies for the AGB stars in this cluster. 

\end{abstract}





\section{Introduction}

Asymptotic Giant Branch (AGB) stars are evolved, low to intermediate mass stars 
(1 $\lesssim$ M$_i$/\Msun\, $\lesssim$ 8). They are characterized by a degenerate inert C-O core 
surrounded by helium and hydrogen nuclear burning shells, burning 
alternatively, forming a double-shell configuration. An intershell 
region rich in helium and carbon exists between the He and H nuclear 
shells. A deep H-rich convective envelope surrounds the double-shell 
configuration \citep[e.g., see the review by][]{herwig05}. During the 
thermally-pulsing AGB (TP-AGB) phase, the He-shell becomes unstable, 
igniting every $\approx$10$^{5}$ years 
or so, with the resulting thermal pulses (TPs) lasting for $\approx$10$^{2}$ years. 
Following TPs mixing episodes can occur. The mixing brings the products of nuclear 
burning of H and He (mostly \iso{4}He and \iso{12}C) from the interior 
of the star to the stellar surface \citep{iben75}. These are referred to as 
third dredge-up (TDU) events. Through the action of repeated TDUs, AGB stars 
evolve from having an oxygen-rich composition (O-rich stars) where C/O $<$ 1 
to a carbon-rich composition (C-rich stars) where C/O $>$ 1. The TDU can 
also result in AGB stars with enhanced s-process elements in their spectra 
\citep[see][]{busso99}. Thus, thermal pulses lead to changes in the surface abundances of 
AGB stars, making them chemically very different compared to their 
less evolved counterparts. The end of the AGB phase is reached when 
the superwind mass loss (up to $\sim$ 10$^{-4}$\Msun\,yr$^{-1}$) 
reduces the hydrogen-rich envelope to small values 
($\lesssim$10$^{-3}$\Msun), with the ejected matter diffusing 
into the inter-stellar medium \citep{bloecker01}.

Star clusters are ideal sites to test theories of stellar evolution. 
They contain stars of similar age and metallicity. 
Star clusters in the Magellanic Clouds (MCs) prove very valuable 
in comparison to the star clusters in our Galaxy as they span a wide 
range of age, which enables us to study the evolution of stars of various 
masses \citep[e.g.][etc]{girardi09,mackey08,milone09,mucciarelli06}. 
The MCs house a large population of rich, intermediate age clusters which 
are useful for studying the the short-lived later stages of stellar 
evolution, especially for stars with masses around 1.5 to 2.5\Msun\,
\citep{girardi95,wood94}. Since we know the distance to 
the MCs accurately, the luminosities of AGB stars can be determined, 
in contrast to AGB stars in the Galaxy which occur mostly in field 
populations. Further, the intermediate-age clusters in the MCs 
demonstrate the TDU brilliantly \citep{bessell83,frogel90} and are 
useful probes to study the details of the TDU in O-rich stars and C-rich stars.

The objective of our work is to study the evolution and nucleosynthesis 
of AGB stars in the two Large Magellanic Cloud (LMC) clusters, NGC\,1978 and NGC\,1846, and 
in the Small Magellanic Cloud (SMC) cluster NGC\,419. These three 
clusters are ideal test beds for evolution and nucleosynthesis 
studies owing to the availability of accurate estimates of AGB structural 
parameters such as pulsation masses, effective temperatures 
($T_{\rm{eff}}$), and luminosity (Kamath et al. 2010 for NGC\,1978 and 
NGC\,419, and Lebzelter \& Wood 2007 for NGC\,1846). Abundance studies 
and attempts to explain the observed C and O abundances for NGC\,1978 and the 
C, O, and F abundances for NGC\,1846 have been carried out by \citet{lederer09b} and 
\citet{lebzelter08}, respectively. It was found that the derived 
C/O and \iso{12}C/\iso{13}C ratios for the two clusters showed 
very different results considering that the two clusters have 
similar AGB masses and metallicities. For NGC\,1978, no 
theoretical scheme was established that could satisfactorily reproduce 
the derived chemical abundance pattern. Further, it was found that 
the M-stars in NGC\,1846 showed 
a very rapid increase in the observed [F/Fe] versus C/O ratios compared to 
the predictions \citep{lebzelter08}. For NGC\,419, no detailed observational 
information on abundances exist. 

In our study, we compute new evolution 
models with updated opacities for the AGB 
stars in these clusters. We constrain our models based on the accurate 
observational parameters. This allows us to reproduce observables 
such as the giant branch temperatures, the oxygen to carbon transition 
luminosity (where C/O $\approx$ 1), and the AGB-tip luminosity. We 
then use these stellar evolutionary sequences in a post-processing 
code to study detailed nucleosynthesis and try to reproduce the observed abundances of 
the AGB stars in our target clusters. We estimate the effects of 
extra-mixing processes on the surface abundances, similar to those 
attempted in previous studies by \citet{lebzelter08}, \citet{lederer09b}, 
and \cite{karakas10b}. We also simulate the effects of an intershell 
enhanced in both \iso{12}C and \iso{16}O to explain the unusual 
abundances of the cluster AGB stars. 

 
The outline of this paper is as follows. In Section~\ref{sec:targetclusters} 
we introduce the three target clusters and supply existing 
abundance information for them. In  Section~\ref{sec:stellarmodels} 
we present a general overview on the numerical method involved in 
the stellar evolution and nucleosynthesis models that we use in our 
work. Here we also discuss the uncertainties in stellar models and 
details on the updated stellar evolution code. In Section~\ref{sec:results} 
we present details of the individual models for each cluster and we 
discuss our results. Finally, in Section~\ref{sec:DnC} we summarize our 
work and provide some concluding remarks. 

\section{Target Clusters}
\label{sec:targetclusters}

\subsection{NGC\,1978}
\label{sec:ngc1978}

The rich intermediate-age cluster NGC\,1978 is an interesting candidate in the LMC. 
This cluster houses both M-type and C-type stars 
\citep{lloyd80,lederer09b,kamath10}. From pulsation analysis, the cluster 
is found to have red variables early on the AGB with a mass of 1.55 $\pm$ 0.1\,\Msun\, 
\citep{kamath10}. This study also showed that the highly evolved AGB stars 
have had a substantial amount of mass loss. 
NGC\,1978 is the only cluster in the LMC with a known mid-IR 
source \citep{tanabe98} which has a large infra-red excess, indicative of 
a very large mass-loss rate.

The properties of the cluster are given in \citet{kamath10}. The metallicity 
estimates are mostly in the range [Fe/H] = $-$0.37 dex to $-$0.42 dex with no 
evidence for any $\alpha$-enhancement \citep{mucciarelli08} and the age 
is estimated to be $\tau = 1.9\pm0.1$ Gyr. The age and metallicity estimates 
lead to an initial mass for the current AGB stars of 1.54 to 1.62\,\Msun, consistent 
with the direct pulsation mass determinations for early-AGB stars of 
1.55 $\pm$ 0.1\,\Msun.

\citet{lederer09b} derived C/O and \iso{12}C/\iso{13}C ratios for nine AGB stars 
in this cluster. They found that, for the M-stars in their sample, the C/O 
ratio ranges from 0.13 to 0.18 with a typical uncertainty of $\pm$0.05. 
The \iso{12}C/\iso{13}C ratio values range between 9 and 16 with an 
uncertainty of up to $\pm$4. They attributed these low 
C/O and \iso{12}C/\iso{13}C ratios for the M-stars in the cluster to 
the fact that the M-stars have not undergone any TDU. 
Based on a sample of four C-stars, for which reliable abundances could 
be established for only two, they found that the C/O ratio was around 1.35 
with an uncertainty of up to $\pm$0.10 and the corresponding 
isotopic carbon ratio was about 150 to 175 with significantly larger errors 
($\approx$ $\pm25$ to $\pm50$). 

\subsection{NGC\,1846}
\label{sec:ngc1846}

NGC\,1846 is a rich intermediate-age LMC cluster known to have a very 
interesting CMD. Using HST observations, \citet{mackey07} identified the presence 
of two distinct main-sequence turn-offs that are clearly associated with 
the cluster. They are the result of the presence of two separate stellar populations 
with the same metallicities 
but different ages of $\tau_{\rm u}$ = 1.5 Gyr for the upper turn-off 
and $\tau_{\rm l}$ = 1.8 Gyr for the lower 
turn-off. Recent work by \citet{goudfrooij09} using HST data 
confirmed the double main sequence turn-off feature as well as identified 
the presence of a RGB bump, a feature that is also found in NGC\,1978. 
They found that age values of 1.7 Gyr and 2.0 Gyr for the upper and 
lower turn-off's respectively, AGB star masses between 1.70 and 1.77\,\Msun, and 
best-fit abundances of [Fe/H] = $-$0.50 with [$\alpha$/Fe] = 0.20. 
\citet{lebzelter07} derived pulsation masses for the AGB stars in NGC\,1846 of 
1.8\Msun, which corresponds to a cluster age of 1.9 Gyr.

\citet{lloyd80} identified a large number of M-stars, 
C-stars, and S-stars in NGC\,1846. The cluster AGB stars do not show significant 
mass loss along the AGB \citep{lebzelter07}. Further, no stars with a high 
mid-IR excess have been found in NGC\,1846 \citep{tanabe98}. 
\citet{lebzelter08} determined the C/O, \iso{12}C/\iso{13}C and the [F/Fe] 
ratios for a small sample of AGB stars in this cluster. For the M stars, 
they determined C/O ratios between 0.2 and 0.65 with an uncertainty 
of up to 0.1 dex. Carbon isotopic ratios varying between 12 and 60 were found for the 
sample of M-stars. For the C-stars in the sample, they derived a C/O value of around 
1.8 and an isotopic carbon ratio of about 60. For the M-stars in 
their sample, they also measured the change of fluorine 
abundance along the AGB using the blended HF line and found a clear 
increase in the F abundance with luminosity with the [F/Fe] values ranging 
between $-$0.71 to 0.40. 

 \subsection{NGC\,419}
\label{sec:ngc419}

NGC\,419 is a populous intermediate-age cluster in the SMC. This cluster houses 
a large population of AGB stars, many of which are C-stars 
\citep{frogel90,mucciarelli08,kamath10}. The properties of this cluster are given in 
\citet{kamath10}. The metallicity is estimated to be around [Fe/H] = $-$0.7 dex and the 
age estimates are mostly in the range 1.2 to 1.6 Gyr. Using the variability of 
stars in this cluster, \citet{kamath10} derived pulsation masses of 
1.87 $\pm$ 0.1\,\Msun\, early on the AGB. The AGB stars in this cluster show significant mass loss along 
the AGB which agrees well with the existence of a mid-IR source detected by the 
ISOCAM survey \citep{tanabe98}.

\section{The Numerical Method}
\label{sec:stellarmodels}

We calculate the stellar evolution and nucleosynthesis in two steps. First, we use 
the stellar evolution code to follow the evolution of the stellar structure and abundances important 
for stellar evolution (H, \iso{3}He, \iso{4}He, \iso{12}C, \iso{14}N, and \iso{16}O) 
from the zero-aged main sequence (ZAMS) to the end of the TP-AGB phase (see Section~\ref{sec:evomods}). 
Then we perform detailed nucleosynthesis calculations (see Section~\ref{sec:nucmods}). 
The numerical method and the procedure used to compute the models 
have been previously described in detail by \citet{karakas02},
\citet{lugaro04}, \citet{karakas10a}, and references therein. Here, we summarize the 
essential details relevant to our study.

\subsection{Stellar Evolution models}
\label{sec:evomods}

The stellar evolution is calculated using an updated version of 
the Mount Stromlo Stellar Evolution Code 
\citep{wood81,lattanzio86,frost96,karakas07b}. 
The masses and compositions of the stars evolved are listed in Section~\ref{sec:results}.
Low-mass stellar models are affected by many uncertainties, the most
important of which are listed below.
With our improved models we aim to constrain these uncertainties. 

\subsubsection{Mass-loss}

Dealing with the extent and temporal distribution of 
mass-loss in AGB stars is a major uncertainty in stellar modeling. 
Model calculations use simple parametrized formulae which are supposed to 
be an average of what is observed. In the existing stellar evolution code, 
the \citet{reimers75} mass-loss prescription is used on the RGB with 
$\eta = 0.4$, and the \citet{vw93} mass-loss prescription is used on the AGB. 
We note that the latter prescription was derived from a sample 
containing both O-rich and C-rich stars so that it should be valid for 
our modeling. Furthermore, at higher mass-loss rates, direct measurements 
of the mass-loss rates for MC AGB stars agrees reasonably well with the 
\citet{vw93} mass-loss rate \citep{wood07b}.
In our study to reproduce the AGB-tip luminosity, we found that the 
\citet{vw93} mass-loss prescription produced AGB-tip luminosities that were 
about 0.2 mag too faint. As a simple correction for this, we modify the \citet{vw93} 
mass-loss prescription and invoke the super-wind phase at a later 
stage during the AGB evolution. This is done by effectively increasing the pulsation 
period at which the super-wind starts from 500 days (as found in Vassiliadis \& Wood 1993) 
to $\approx 710 - 790$ days (specifically, the term '$P$' in 
Equation 2 of \citet{vw93} is replaced by '$P$ $-$ 210' and '$P$ $-$ 290', respectively,
depending on the mass and metallicity combination).
The period values at which the superwind mass-loss rate begins 
(i.e., where the mass-loss rate given by Equation 1 of \citet{vw93} matches 
the value given by the modified Equation 2) for each structure model is 
listed in Table~\ref{table1}.

It is worth commenting on the need to increase the period for 
the onset of the
superwind. The basic reason for this is that in these models with
new C-rich opacities, the AGB stars become considerably cooler when
C/O exceeds unity.  This means that the radius and hence computed
pulsation period increases significantly causing the superwind
to occur at lower luminosities than in O-rich stars or in past models 
without C-rich opacities.  Observationally, as noted above, the
O-rich and C-rich stars in the sample of \citet{vw93} seem to fall
on the same curve of rising mass loss with pulsation period.  Also,
it appears that the $K - \log P$ and $M_{\rm bol} - \log P$ relations are 
indistinguishable, at least while the stars are optically visible 
\citep{feast89}. There seems to be an
inconsistency between the observed $\dot{M} - P$ relation and that
predicted by the stellar models with the new C-rich opacities.  
Our adjustment of the onset period for the superwind is one way
to bring the model mass loss rates and periods back to those that seem to 
apply observationally.  Perhaps new pulsation models with C-rich opacities
will resolve this problem.

\subsubsection{Convection}
\label{sec:Non-std-intershell}

One of the biggest uncertainties in stellar models is the 
treatment of convection. We use the mixing length (MLT) theory for 
convective regions and we set the mixing length parameter $\alpha = \ell/H_{\rm P}$ 
in the models by matching the observed
RGB and E-AGB $T_{\rm{eff}}$, taking observed $T_{\rm{eff}}$ 
values from \citet{kamath10}. 
We then keep the mixing length parameter a constant.

Previous studies by \citet{straniero97}, \citet{karakas02}, 
\citet{stancliffe04b}, and \citet{karakas10b} show that 
different stellar evolution codes predict 
different TDU efficiencies, with some codes predicting no TDU
if overshoot is not included \citep[e.g.,][]{mowlavi99a}. Furthermore, 
low-mass (M $\leq$ 2\,\Msun) AGB models show little or no TDU \citep{karakas02}. 
In our models, to alter the extent of TDU, we include 
convective overshoot by extending the position of the 
base of the convective envelope downward by $N_{\rm ov}$ pressure scale heights
\citep{karakas10b}. We include convective overshoot at the base of the envelope 
at all times during the AGB. By changing $N_{\rm ov}$ we alter the amount of 
carbon that is dredged-up to the outer layers of the star such that the M/C 
transition takes place near the observed $M_{\rm bol}$ \citep{lebzelter07,kamath10}.
We note that we define the M/C transition luminosity as the luminosity 
of the brightest M-star as there is an overlap in the luminosity of M and C-stars 
due to luminosity variations over a thermal pulse cycle. 
The values of $\alpha$ and $N_{\rm ov}$ that we use 
for the models of the cluster AGB stars are listed in Table~\ref{table1}.

During thermal pulses, adding convective overshoot to the base of the 
intershell convection will cause this convection to penetrate further 
into the C-O core, resulting in intershell 
abundances that are different from models with no overshoot \citep[e.g.,][]{herwig00}. 
The subclass of PG1159 post-AGB stars are H-deficient and show 
He-intershell material at their surface \citep{werner09}. 
In these stars, C abundances vary from 15--60\% (by mass) and O from 2--20\%.
These abundances are in direct contrast to standard intershell
compositions that give 25\% C and $\lesssim 2$\% O \citep[e.g.,][]{boothroyd88c}. 
The theoretical models of \citet{herwig00} include the effect of 
diffusive convective overshoot into the C-O core during a TP. This increases the 
C and O intershell abundances well above the value found in standard models
but consistent with PG 1159 star abundances.
Though we do not include this in our evolution models, in our nucleosynthesis study 
for AGB stars in NGC\,1978 (refer Section~\ref{sec:1978results}), we synthetically 
estimate the effects of a non-standard intershell which is enhanced in 
both \iso{12}C and \iso{16}O . We employ methods similar to those 
described in \citet{karakas10b}.

\subsubsection{Opacities}

The outer layers of AGB stars become cool, 
allowing molecules to form. Therefore an accurate treatment 
of molecular opacities is needed. In our version of the stellar evolution code 
we have utilized new opacity tables. At low temperatures we have used the 
Rosseland mean opacities computed using AESOPUS \citep{marigo09} with  
\citet{lodders03} solar abundances as the reference solar composition. 
Using AESOPUS we have the option of incorporating 
scaled-solar, $\alpha$-enhanced, and carbon-depleted opacity tables. We 
have generated tables based on the initial composition we require for each of the clusters. 
We also utilize the OPAL radiative opacity tables of \citet{Iglesias96} 
updated to use a \citet{lodders03} solar abundance distribution of elements from C to Fe. 
To maintain consistency between the abundances of the low-temperature tables and 
the high-temperature OPAL tables we use OPAL tables appropriate for each of the compositions 
(scaled-solar, $\alpha$-enhanced, and carbon-depleted).

\subsection{The nucleosynthesis models}
\label{sec:nucmods}

The evolution code only includes the species (H, \iso{3}He, \iso{4}He, 
\iso{12}C, \iso{14}N, and \iso{16}O) that are relevant to 
the major energy generating reactions and this forms the basis for the structure of 
the model star. In order to try and explain the elemental and isotopic abundance patterns 
observed in stars we need to 
include more nuclear species. We use a post-processing nucleosynthesis code for this purpose. 
This code needs as input from the stellar evolution code variables such as temperature, 
density, and convective boundaries as a function of time and mass fraction. The code then re-calculates 
the abundance changes as a function of mass and time using a nuclear network 
which contains 77 species (from hydrogen to sulphur, along with a small group of
iron-peak elements) and time dependant diffusive mixing for all 
convective zones \citep{cannon93}. The overshoot regions at the bottom of the convective envelope 
in the evolution code are not treated as convective in the nucleosynthesis code, i.e., 
convection and mixing are assumed to stop at the Schwarzschild boundary. The code also 
requires input physics such as reaction rates and initial abundances. 
Most of the 589 reaction rates are taken from the JINA REACLIB data base \citep{cyburt10}.  
Details on the updated reaction rates that we use can be found in \citet{karakas10a}. 
For the three clusters, we assume the initial abundances based on the 
existing observational information on their abundances. Details on the 
input abundances used for each model are discussed in 
Section~\ref{sec:results}. We take our reference solar composition from \citet{lodders03}. 

An additional feature of the post-processing nucleosynthesis code which 
is not in the evolution code is that we can estimate the 
effect of extra-mixing. Observations 
of low-mass red giant stars ($M \lesssim 2\Msun$ near the RGB-tip)
reveal \iso{12}C/\iso{13}C ratios of $\sim 10$ and C/N 
$\sim 1.0$ \citep{gilroy89}. These ratios are lower than predicted
by standard stellar evolution models, which give
\iso{12}C/\iso{13}C $\sim 20$ and \iso{12}C/\iso{14}N $\sim 1.5$.
\citep[e.g.][]{charbonnel94}. These trends are also seen
in globular clusters where there is an anti-correlation of C abundance with 
luminosity \citep[e.g., in M3;][]{gsmith02}. Together these observations
indicate the occurrence of non-convective mixing processes on the 
giant branch. Mechanisms proposed to account for this 
extra-mixing \citep[see][]{herwig06} include rotational mixing 
\citep{charbonnel98}, gravity waves \citep{denissenkov00}, thermohaline mixing 
\citep{eggleton08,charbonnel07,stancliffe09,stancliffe10}, and magnetic 
fields \citep{nordhaus08,busso07b,palmerini09}. 
To simulate the effect of extra-mixing on the RGB, we take the envelope 
composition at the tip of the giant branch from the nucleosynthesis
calculations and alter it such that the 
\iso{12}C/\iso{13}C ratio equals the observed M-star \iso{12}C/\iso{13}C ratio. 
This is done by decreasing the \iso{12}C abundance, and increasing the
abundances of \iso{13}C and \iso{14}N in the entire convective envelope 
according to changes expected for CN cycling, i.e., the total number of 
\iso{12}C, \iso{13}C, and \iso{14}N nuclei are conserved. The modified 
models are subsequently evolved through core-helium burning and on the AGB.
An extensive explanation on this procedure can be found in 
\citet{karakas10b}.

\section{Models and Results} 
\label{sec:results}

In this section we present the stellar evolution and nucleosynthesis models 
for the AGB stars in our target clusters. The parameters used to
construct the stellar models are listed in Table~\ref{table1}. These include 
the initial mass and metallicity, the initial abundance pattern, the 
$\log$ $T_{\rm eff}$ for the M-stars on the AGB at $M_{\rm bol} = -4.00$ 
($\log$ $T_{\rm eff,4}$), 
the mixing-length parameter ($\alpha$) used to fit the giant branch $T_{\rm eff}$, 
the mass on the early AGB ($M_{\rm e-agb}$), the amount of
convective overshoot required to fit the M/C transition luminosity 
($N_{\rm ov}$), the bolometric luminosity of the model star at the M/C
transition ($M_{\rm bol}^{\rm M/C}$), the pulsation period ($P$) 
where the superwind mass-loss rate begins, and the predicted bolometric luminosity at the tip
of the AGB ($M_{\rm bol}^{\rm agb-tip}$).

We use starting compositions with a variety of C/O ratios in our evolution 
calculations. These C/O ratios were chosen to broadly match the C/O ratios 
observed for the cluster M-stars which lie on the early-AGB. 
For NGC\,419 we construct models using a scaled-solar abundance pattern as there 
are no observed C/O ratios. However for NGC\,1978 and NGC\,1846 we 
experiment with three different initial compositions as there are observed 
C/O ratios: a scaled-solar composition; a carbon-depleted composition 
where [C/Fe] = $-$0.25 dex with scaled-solar 
values for the other elements (hereafter, the carbon depleted model); and an
$\alpha$-enhanced composition where [$\alpha$/Fe] = $+$0.20 dex while for all the other 
elements we assume scaled-solar values (hereafter, the $\alpha$-enhanced model). 
For all the above mixtures we use the \citet{lodders03} reference solar 
composition. 

In Table~\ref{table2} we present some of the details of the stellar
structure models. The first row for each cluster gives the initial mass and metallicity. 
Then we include for each model the number of TPs computed, 
the H-exhausted core mass (hereafter core mass) at which the TDU 
begins ($M_{\rm c}^{\rm min}$), 
the maximum TDU efficiency ($\lambda_{\rm max}$, where 
$\lambda$ = $\triangle$$M_{\rm dredge}$/$\triangle$$M_{\rm h}$, 
$\triangle$$M_{\rm h}$ being the amount by which the core mass 
has grown between the present and previous TPs), and the average value for the 
TDU efficiency ($\lambda_{\rm avg}$). We also include
the total amount of mass dredged into the envelope during the AGB lifetime 
($M_{\rm dredge}$), the maximum He-shell temperature 
($T_{\rm Heshell}^{\rm max}$) which is usually taken from
the final TP, the core mass and the total mass at the
final (the last computed) timestep ($M_{\rm c}$(f) and $M_{\rm tot}$(f), respectively), and 
the final interpulse period ($\tau_{\rm ip}$(f)).

\subsection{Evolution and Nucleosynthesis model results for NGC\,1978}
\label{sec:1978results}

\begin{figure}
\begin{center}
\includegraphics[width=15cm,angle=0]{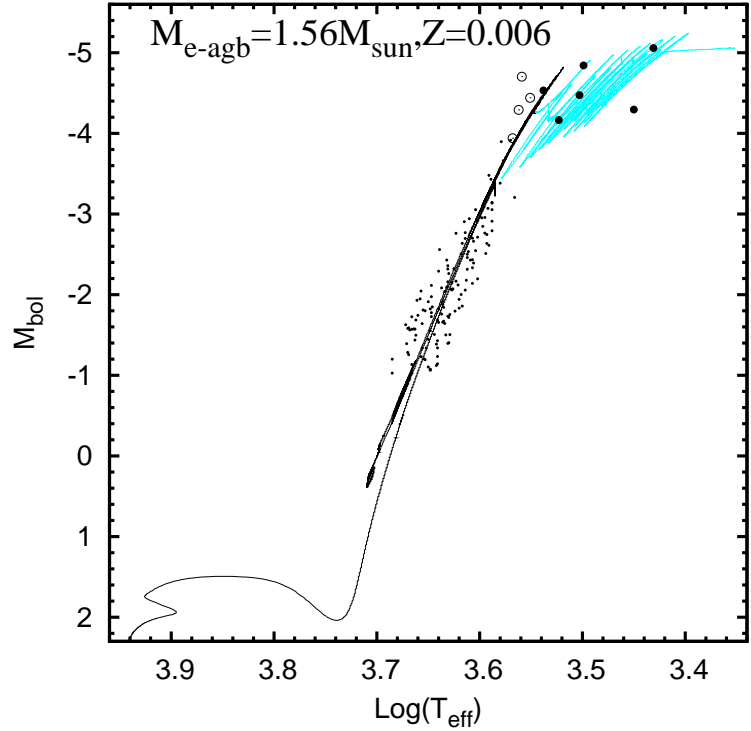} 
\caption{The HR diagram for the carbon-depleted model for NGC\,1978. 
The cyan/grey lines mark the M/C transition and indicates that the star is C-rich. 
The black open circles represent the observed positions of the M-stars in NGC\,1978, the black 
filled circles represent the observed positions of the C-stars in NGC\,1978, and 
the smaller black dots represent the non-variable red-giants in the cluster \citep{kamath10}. 
Note that the low luminosity of the C-star indicates that the star is in a post-flash 
luminosity dip.}
\label{n1978}
\end{center}
\end{figure}

Based on the estimated AGB pulsation mass of $\sim$1.55\,\Msun\, for the cluster AGB 
stars (see Section~\ref{sec:ngc1978}) we construct evolutionary sequences
starting from the ZAMS having an initial 
mass of 1.63\,\Msun\,with $Z$ = 0.006 (similar to the observed global 
metallicity for this cluster, [Fe/H] $\approx -0.4$) and $Y$ = 0.25. 
We construct stellar evolution models based on three 
initial compositions: scaled-solar, carbon-depleted, and $\alpha$-enhanced.
To match the composition of the lowest M-star C/O ratio, we assume 
that the $\alpha$-enhanced model also has a slight carbon depletion of 
[C/Fe] = $-$0.05 dex. Figure~\ref{n1978} shows the
theoretical Hertzsprung-Russell (HR) diagram for the carbon-depleted model as an example. We also 
overplot the AGB variables and the non-variable red-giants in NGC\,1978 
\citep{kamath10}. We find that the evolutionary track is in good agreement with the 
observed star positions. The other models corresponding to the other two initial compositions 
have similar evolutionary tracks.
The values for the  mixing length parameter $(\alpha$) 
are the same for all three compositions at 1.90. However the amount of 
overshoot varies slightly ($N_{\rm ov}$ $\approx$ 2.45 $-$ 3.00), with the
$\alpha$-enhanced model requiring the most overshoot and the 
carbon-depleted model requiring slightly less overshoot 
than the scaled-solar model (refer to Table~\ref{table1}). 
The amount of overshoot used in each model series results in 
predicted M/C transition $M_{\rm bol}$ values that give a good match to the observed 
M/C transition bolometric luminosity of $M_{\rm bol} = -4.5$ \citep{kamath10}.
Observations from \citet{kamath10} also indicate that the most luminous cluster AGB star has an 
$M_{\rm bol}= -5.06$. To match this luminosity, the super-wind phase for the three models 
start at a pulsation period of 790 days.

\begin{figure}
\begin{center}
\includegraphics[width=12.5cm,angle=0]{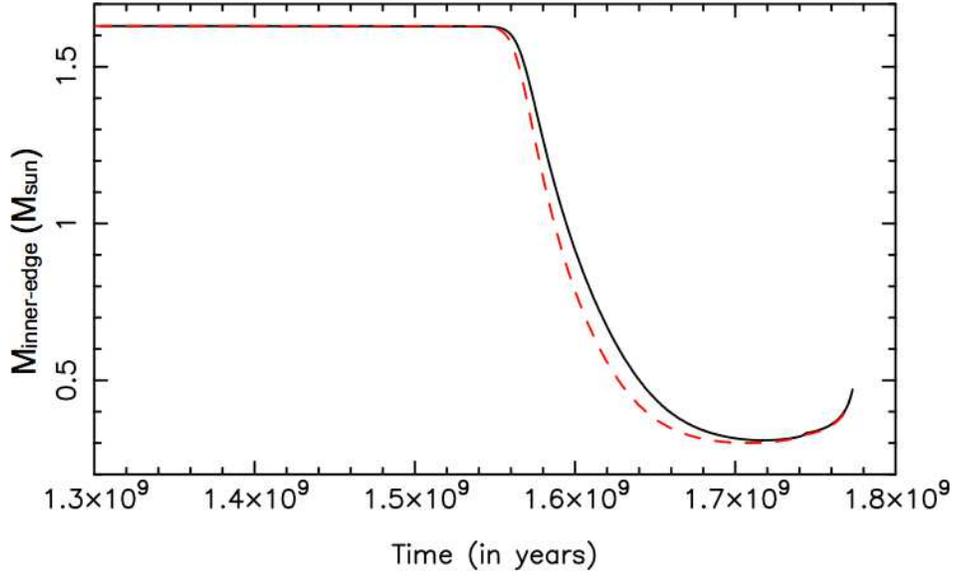} 
\caption{Mass interior to the inner edge of the convective envelope, 
showing the depth of FDU for the scaled-solar NGC\,1978 model. The black solid
  line represents the standard model without overshoot, and the red dashed-line 
indicates the case with mild RGB overshoot.}
\label{n1978_ov}
\end{center}
\end{figure}

One interesting feature of NGC\,1978's CMD is the presence of the RGB bump.
This is because the RGB bump luminosity can be used as an extra 
observational constraint on the stellar
structure models. The RGB bump is caused by the H-shell erasing
the abundance discontinuity left by the retreating convective envelope
during the first dredge-up (FDU). In Fig.~\ref{n1978_ov} we show the location (in mass) of 
the inner edge of the convective envelope during the FDU. The black
solid-line shows the scaled-solar model, and this model has a slightly
higher RGB bump bolometric luminosity of $M_{\rm bol}$ = $-$0.53 when compared to the observed 
$M_{\rm bol}$ $\approx -$0.31, calculated from the position of the RGB bump 
at V$_{\rm 555}$ = 19.10 $\pm$ 0.10 \citep{mucciarelli07}. To match
the observed RGB bump luminosity we can apply convective overshoot in
a similar way as done for the AGB. The red dashed-line in  Fig.~\ref{n1978_ov} 
shows the result of one such model with a $N_{\rm ov}^{\rm RGB}$ = 0.30. The figure shows
that the depth of the FDU is not significantly altered, although the FDU begins at an
earlier time and this lowers the RGB bump bolometric luminosity 
to better match the observed values. The structural details (e.g., surface luminosity, 
core mass, effective temperature, etc) were essentially the same at the tip
of the RGB and on the E-AGB in the model with slight overshoot 
compared to the model without. For this reason we will ignore overshoot
on the RGB for the rest of this study.

\begin{figure}
\begin{center}
\includegraphics[width=13cm,angle=0]{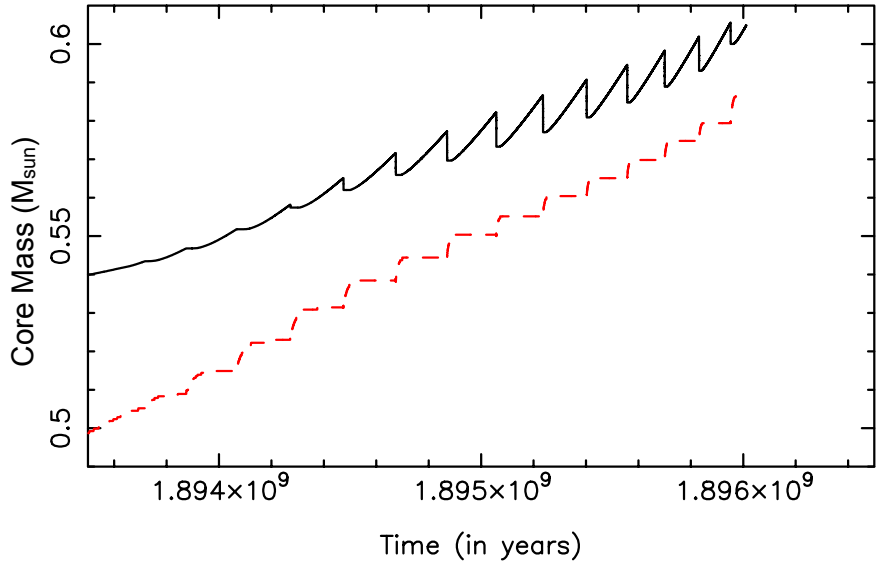} 
\caption{Core masses for the scaled-solar NGC\,1978 model. The black solid
  line shows the variation of the H-exhausted core mass with time, and the red dashed-line 
indicates the time variation of the He-exhausted core mass.}
\label{TDUeff}
\end{center}
\end{figure}

From Table~\ref{table2} we find that all the three models in NGC\,1978 experience $\approx$ 15 
TPs. To reproduce the observed M/C transition luminosity we require very extended 
TDU with $N_{\rm ov} \sim 2.45 - 3.00$. This results in $\lambda_{\rm max}$ of 0.72 $-$ 0.82 
(see Fig.~\ref{TDUeff}) and an average $\lambda$ of 0.58 $-$ 0.65 
for all TPs. When convergence difficulties terminated the evolution 
calculations (at an envelope mass of the order of $\sim$ 0.05 to 0.20\Msun\, for the 
different models computed), the mass-loss rates were so high that we would not 
expect any further TPs or TDU events to occur on the AGB.

We now discuss the results of the post-processing nucleosynthesis calculations. 
In Table~\ref{table3} we include C, O, and F surface abundance information for all 
the stellar models. We provide the initial, post-FDU, and final computed C/O ratios 
(C/O$_{\rm i}$, C/O$_{\rm post-FDU}$, and C/O$_{\rm f}$, respectively), where
all abundance ratios are number fractions. Furthermore we list the initial, post-FDU, and final computed 
\iso{12}C/\iso{13}C ratios 
(\iso{12}C/\iso{13}C$_{\rm i}$, \iso{12}C/\iso{13}C$_{\rm post-FDU}$, and 
\iso{12}C/\iso{13}C$_{\rm f}$, respectively). In the last column we provide
the [F/Fe] ratio at the tip of the AGB where we use the standard notation
[X/Y] = $\log_{\rm 10}$(X/Y)$_{\rm star} - \log_{\rm 10}$(X/Y)$_{\rm \odot}$. 

From Table~\ref{table3} we see that the scaled-solar model
gives a post-FDU C/O = 0.33, significantly higher than the highest M-star
C/O ratio of 0.18. On the other hand the carbon-depleted models all yield a C/O ratio of $\sim$ 0.18, 
within the range of the M-star observations (C/O = 0.13 to 0.18). However, the 
predicted post-FDU \iso{12}C/\iso{13}C ratio does not match the observed range of 
\iso{12}C/\iso{13}C = 9 to 16 for all carbon depleted models. In order to obtain a match, we assume extra-mixing has 
taken place on the RGB (as explained in Section~\ref{sec:nucmods}). We alter the 
envelope composition at the tip of the RGB such that \iso{12}C/\iso{13}C = 13, 
which agrees well with the observed average M-star \iso{12}C/\iso{13}C value of $\approx$ 13. 
We designate AGB evolutionary sequences in which the abundance 
change has been made as extra-mixing sequences. 

\begin{figure}
\begin{center}
\includegraphics[width=14cm,angle=0]{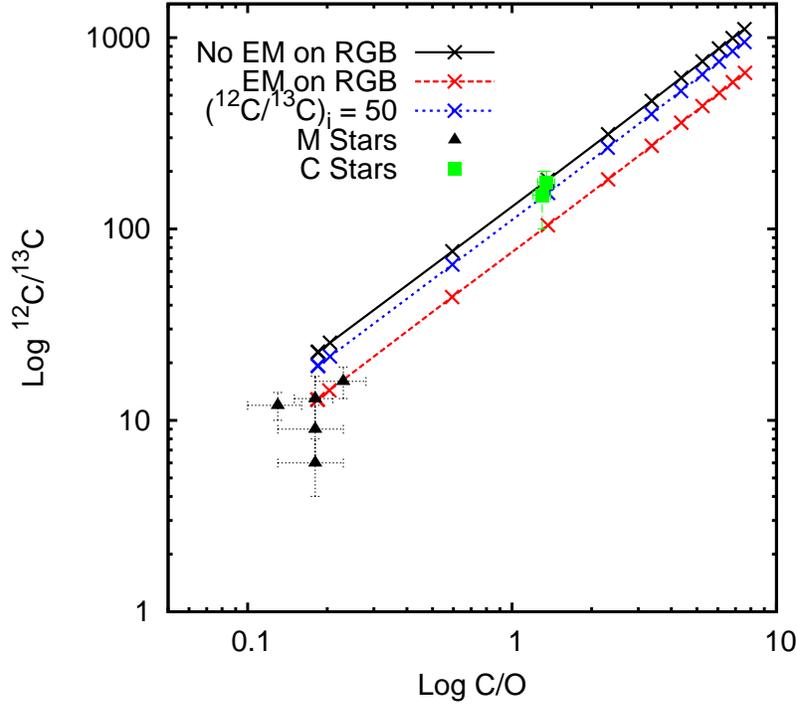} 
\caption{Log C/O versus $\log$ \iso{12}C/\iso{13}C ratios for NGC\,1978's carbon-depleted model. Also
  shown are the observational data for AGB stars in NGC\,1978
  \citep{lederer09b}, where black filled triangles represent the M stars 
  and the green filled squares represent the C stars. The black solid-line
  refers to the model without any extra-mixing, 
the red dashed-line represents the model with extra-mixing on the RGB, and
  the blue dotted-line has an initial \iso{12}C/\iso{13}C = 50. The 
crosses in this plot and subsequent plots denote the envelope
  abundances after each TDU episode.
 }
\label{n1978_nuc_cdep}
\end{center}
\end{figure}

In Fig.~\ref{n1978_nuc_cdep} we show the evolution of the C/O versus
\iso{12}C/\iso{13}C ratios for the carbon-depleted models on a logarithmic scale. Included on the
plot are the observed values for the M and C-type stars in NGC\,1978 from \citet{lederer09b}. 
The black solid-line represents the carbon-depleted model without extra-mixing 
on the RGB and the red dashed-line represents the carbon-depleted model with extra-mixing on the RGB. 
Furthermore, we also perform one nucleosynthesis calculation with a lower
initial \iso{12}C/\iso{13}C = 50 compared to the solar value of 89
\citep[from LMC HII regions, see discussion in][]{wang09}. 
From Fig.~\ref{n1978_nuc_cdep} we see that both the models with no extra-mixing, 
i.e., either with a solar \iso{12}C/\iso{13}C (black solid-line) or with an initial
\iso{12}C/\iso{13}C = 50 (blue dotted-line), have a post-FDU abundance of
\iso{12}C/\iso{13}C which is higher than the observed
M-star value. However, the model with extra-mixing (red dashed-line) has a post-FDU abundance of 
\iso{12}C/\iso{13}C = 12.8 which by construct agrees well with the 
observed average M-star \iso{12}C/\iso{13}C value of $\approx$ 13. 

We also experiment with two $\alpha$-enhanced models, one without extra-mixing on the RGB 
and one with extra-mixing on the RGB (see Table~\ref{table3}), which result in post-FDU C/O 
and \iso{12}C/\iso{13}C ratios similar to the carbon-depleted case. However, NGC\,1978 shows no 
evidence for an initial $\alpha$-enhancement \citep{mucciarelli08}. 
Thus, we favour a model with a carbon-depleted initial composition coupled with 
extra-mixing on the RGB since it accurately reproduces the observed average C and O 
composition of the cluster M-stars. However, this scenario does not adequately explain the cluster C-stars. 
Previously, \citet{lederer09b} attempted to reproduce the observed carbon 
and oxygen abundances for the cluster AGB stars using models with an 
E-AGB mass = 1.55\Msun\, and Z = 0.006 along with moderate extra-mixing 
on the RGB coupled with an [O/Fe] = 0.2 dex. These models were 
able to match the observed average C/O and \iso{12}C/\iso{13}C for the M-stars
but they were unable to reproduce C/O and \iso{12}C/\iso{13}C for the C-stars.  

\begin{figure}
\begin{center}
\includegraphics[width=10cm,angle=270]{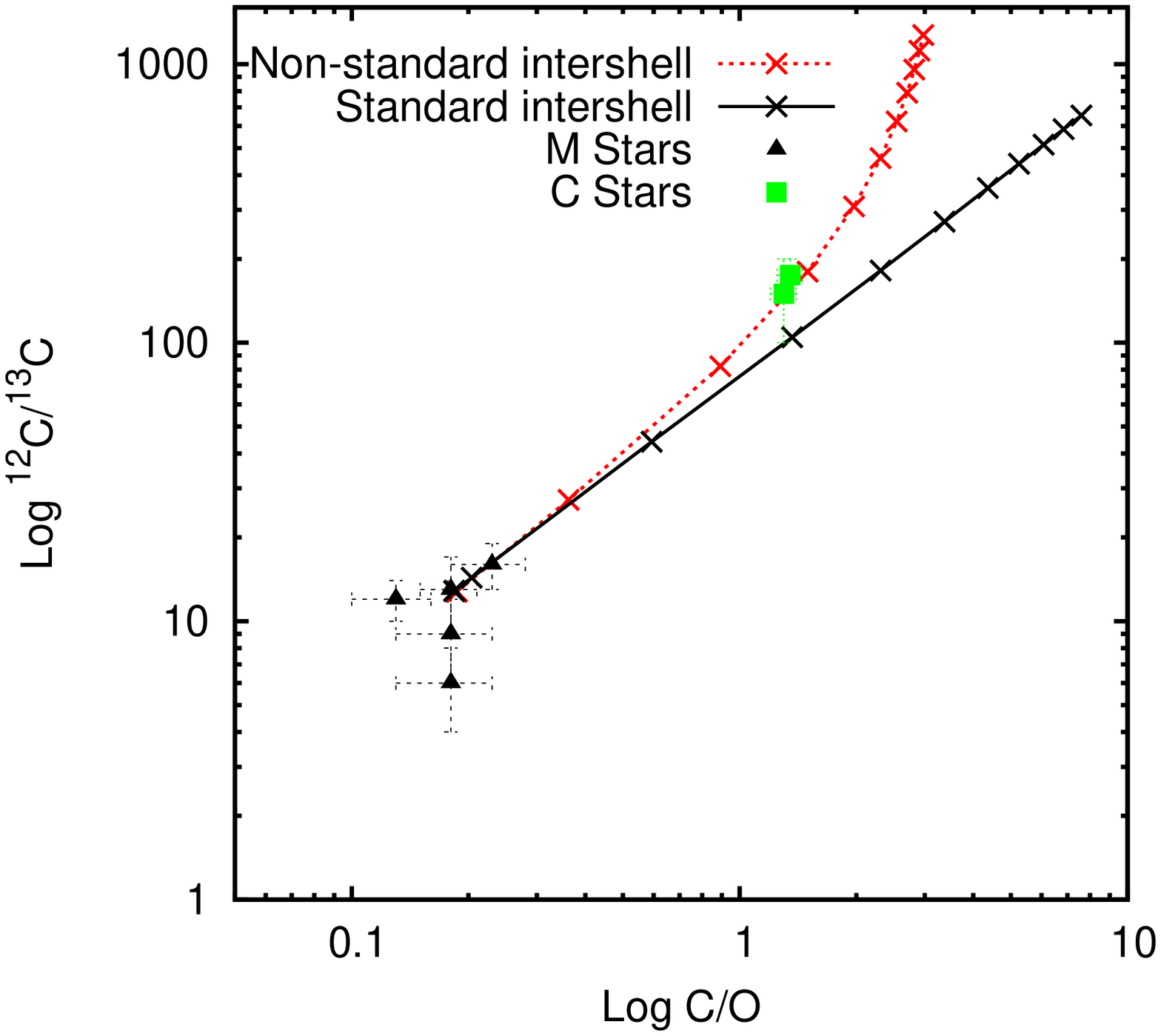} 
\caption{Same as Fig.~\ref{n1978_nuc_cdep} but showing results for NGC\,1978's 
carbon-depleted model with extra-mixing (black solid-line), 
and the synthetic AGB model with an intershell enhanced in carbon and oxygen 
(red dashed-line). In the synthetic model we have assumed intershell mass fractions of 15\%
for O and 40\% for C.}
\label{n1978_nuc_cdep_result}
\end{center}
\end{figure}

The C-stars in NGC\,1978 have large observed carbon isotopic ratios indicating
a strong contribution of \iso{12}C from the He-intershell mixed into the 
envelope by TDU. We note that the models show a slope very close to 1 implying that 
essentially only \iso{12}C increases with time: any substantial change in slope would require 
an additional change to either \iso{13}C or O. One possible 
way to do this would be to have an intershell abundance 
with more \iso{16}O than that predicted by 
standard models as discussed in Section~\ref{sec:Non-std-intershell}. 
In  Figure~\ref{n1978_nuc_cdep_result} we show
a synthetic AGB model prediction (red dashed-line) that results from
increasing the C content to 40\% and the O content of the intershell to 15\% (by mass). 
The standard values for this model are 15\% for C and 0.35\% for O (by mass). 
The artificially enhanced numbers are within the range of C and O abundances
observed for PG 1159 stars \citep{werner06}. With the enhanced intershell C and O values, 
the total elemental C and O surface abundances increase as a result of such mixing by 
factors of $\sim$177 and $\sim$10, respectively, in comparison to a model with a 
standard intershell which results in an increase by factors of $\sim$ 86 and 
$\sim$1.05 in the surface composition. From Figure~\ref{n1978_nuc_cdep_result} 
we see that in NGC\,1978 such a model fits the data for both 
the M-stars and the C-stars very well. Note that the final C/O ratio (C/O $\approx$3.00) is 
much lower than the final C/O ratio of the model with a standard intershell composition (C/O $\approx$ 7.60). 
This implies that some AGB stars may have intershells with higher abundances of 
\iso{16}O and \iso{12}C than those predicted by standard models.

\subsection{Evolution and Nucleosynthesis model results for NGC\,1846} 
\label{sec:1846results}

\begin{figure}
\begin{center}
\includegraphics[width=15cm,angle=0]{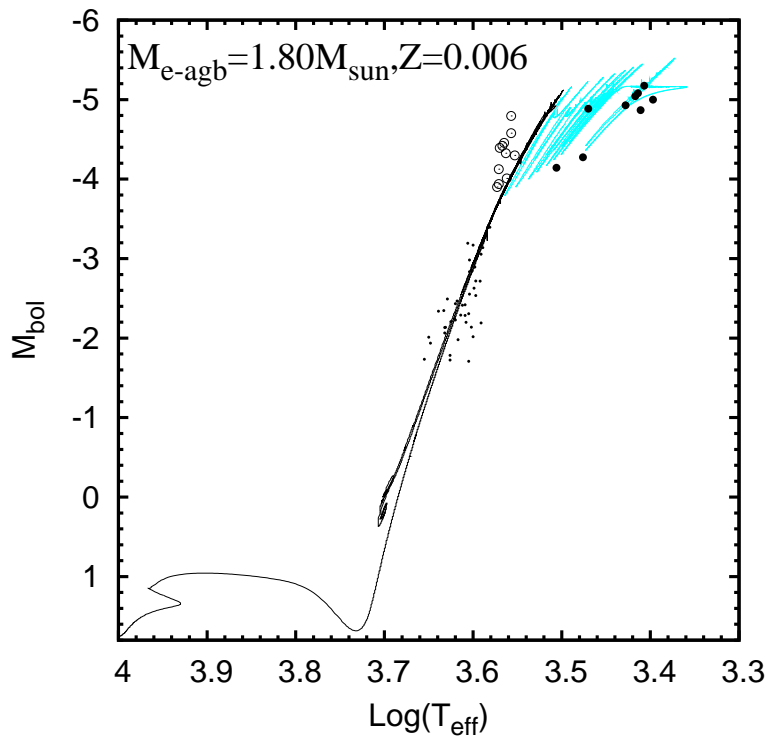} 
\caption{The theoretical HR diagram for the $\alpha$-enhanced model for NGC\,1846. 
The cyan/grey lines indicate that the star is C-rich. The black open 
circles represent the observed positions of the M-stars in NGC\,1846, the black 
filled circles represent the observed positions of the C-stars in the NGC\,1846, and 
the smaller black dots represent the non-variable red-giants in the cluster \citep{lebzelter07}. 
Note that the low luminosity of the C-star indicates that the star is in a post-flash luminosity dip.}
\label{n1846_HR}
\end{center}
\end{figure}

For the AGB stars in NGC\,1846 we construct evolutionary sequences 
starting from the ZAMS having an initial mass of 1.86\Msun\, with Z = 0.006 and Y = 0.25. This 
results in a predicted E-AGB mass of 1.80\Msun\, which is similar to the estimated pulsation 
mass of $\approx$1.80\Msun\, for the cluster AGB stars \citep{lebzelter07}. 
We compute stellar evolution sequences with 
three initial compositions: scaled-solar, carbon-depleted, and $\alpha$-enhanced. 
Figure~\ref{n1846_HR} shows the theoretical HR diagram for 
the $\alpha$-enhanced model as well as the AGB variables and the non-variable red-giants 
in the cluster \citep{lebzelter07}. We find that the observations generally 
match the theoretical evolutionary 
track well although the most luminous M-stars are bluer than predicted by the 
evolutionary tracks. The HR diagrams for the models corresponding to the other two abundances mixes are similar. 
In Table~\ref{table1} we list the input parameters required for the three sequences. 
We find that the value for $\alpha$ is the same for all three compositions at 1.74. 
Overshoot parameters of $N_{\rm ov}$ $\approx$ 1.05 are needed to reproduce 
the observed M/C transition luminosity of $M_{\rm bol} = -4.78$ \citep{lebzelter07} for the 
scaled-solar and the carbon-depleted model while the $\alpha$-enhanced 
model requires slightly more overshoot ($N_{\rm ov}$ $\approx$ 1.41). 
The AGB-tip bolometric luminosity for NGC\,1846 is 
$\approx -5.18$ \citep{lebzelter07}. To match this luminosity the superwind phase 
had to start at a pulsation period of $\approx 710$ days.

The models for NGC\,1846 experience around 16 TPs (see Table~\ref{table2}). 
The NGC\,1846 models have 
slightly less efficient TDU ($\lambda_{\rm avg} \sim 0.46 $) when the 
observed M/C transition luminosity is reproduced, when compared to the NGC\,1978 models 
($\lambda_{\rm avg} \sim 0.62 $) consistent with the smaller amount of overshoot 
($N_{\rm ov}$ $\approx$ 1.05 $-$ 1.41 compared to $\approx$ 2.54 $-$ 3.00 in NGC\,1978). For the NGC\,1846 
model sequences, the final envelope masses when convergence difficulties terminate the evolution 
calculations range between $\approx$ 0.075 to 0.20\Msun. We would not 
expect any further TPs or TDU episodes.

\begin{figure}
\begin{center}
\includegraphics[width=10cm,angle=270]{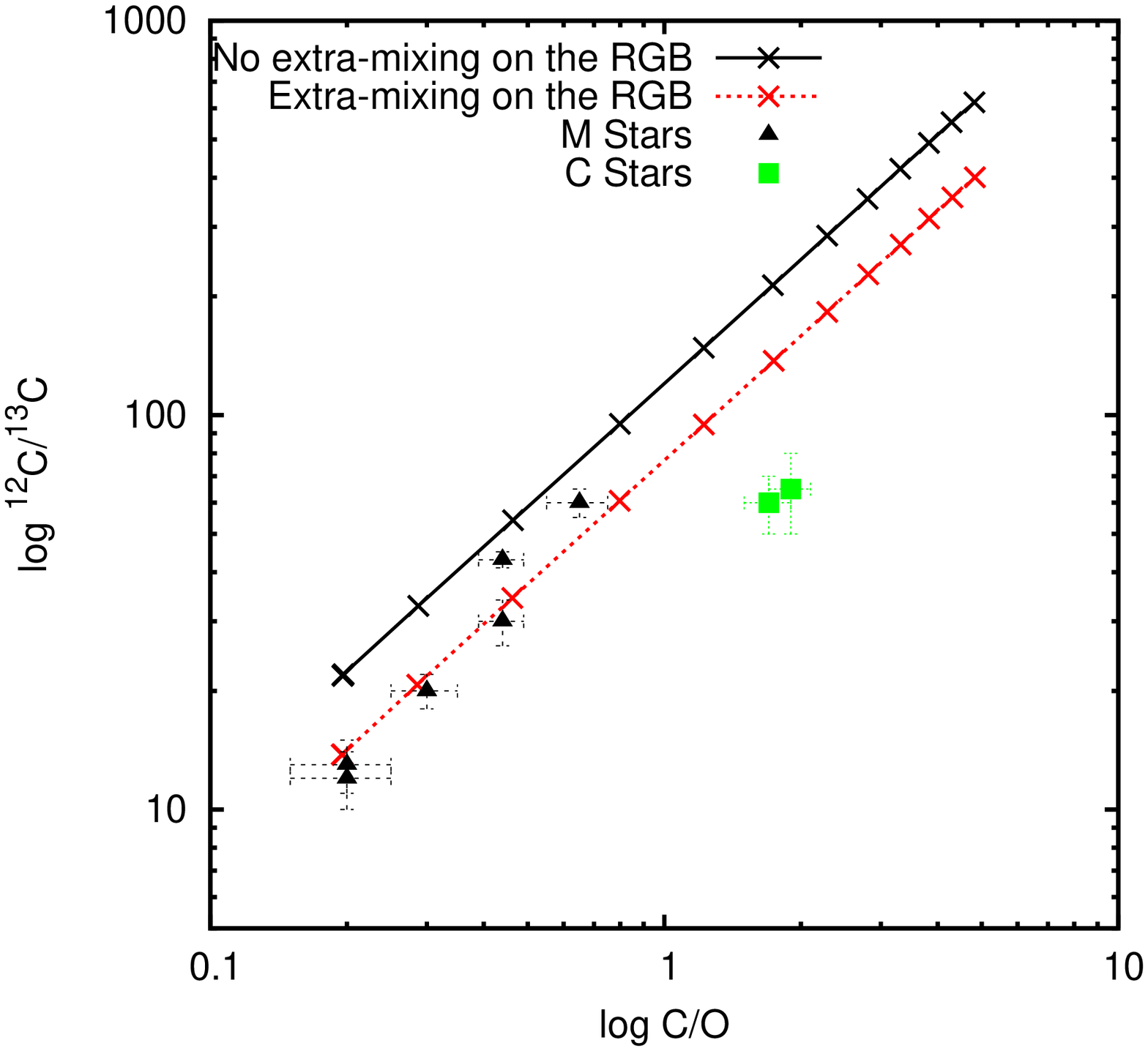} 
\caption{Log C/O versus $\log$ \iso{12}C/\iso{13}C ratios for 
NGC\,1846's $\alpha$-enhanced model. Also shown are the observational data for AGB stars in NGC\,1846 
\citep{lebzelter08}, where black filled triangles represent the M-stars 
and the green filled squares represent the C-stars. The black solid-line 
refers to the $\alpha$-enhanced model without any extra-mixing, 
the red dashed-line represents the $\alpha$-enhanced model with extra-mixing on the RGB.
}
\label{n1846_nuc_aenh}
\end{center}
\end{figure}

Observations indicate that NGC\,1846 shows a mild $\alpha$-enhancement
\citep{mucciarelli08}.  Thus we choose the $\alpha$-enhanced model as our favored model. 
Post-processing nucleosynthesis calculations for the 
$\alpha$-enhanced model  start with an initial 
surface value of C/O = 0.31 and \iso{12}C/\iso{13}C = 89, and yield post-FDU values
of C/O = 0.19 and \iso{12}C/\iso{13}C = 22. We then alter the envelope 
\iso{12}C/\iso{13}C composition such that \iso{12}C/\iso{13}C = 14 at the tip of the 
RGB, in a similar manner as for NGC\,1978 (Table~\ref{table3}).
The M-stars in the cluster show a spread in C/O ratios ranging from 
0.20 to 0.65 and \iso{12}C/\iso{13}C ratios between 12 and 60. 
In Figure~\ref{n1846_nuc_aenh} we show results from the $\alpha$-enhanced
models with extra-mixing on the RGB (red dotted-line) and 
without (black solid-line). From Figure~\ref{n1846_nuc_aenh} it is clear that an
$\alpha$-enhancement combined with extra-mixing on the RGB gives a good match
to both the C/O and \iso{12}C/\iso{13}C ratios of the cluster M-stars. 
However, the subsequent AGB evolution from the M-stars does not 
explain the observed abundances for the cluster's C-stars.

\begin{figure}
\begin{center}
\includegraphics[width=10cm,angle=270]{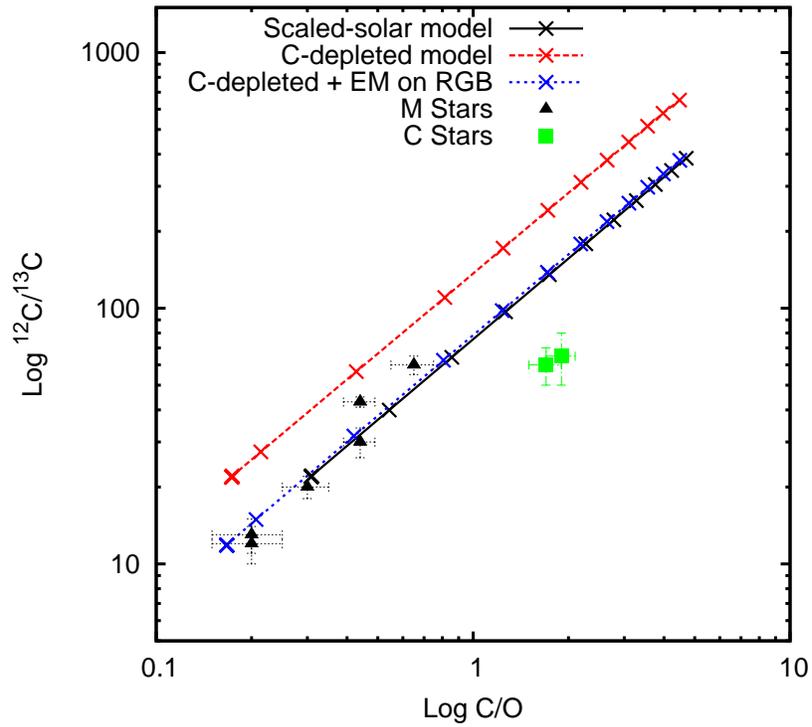} 
\caption{The same as Fig.~\ref{n1846_nuc_aenh} for NGC\,1846's additional model sequences. 
The black solid-line represents the scaled-solar model without extra-mixing on the RGB. 
The red dashed-line represents the carbon-depleted model without extra-mixing on the RGB and the 
blue dotted-line represents the carbon-depleted model with extra-mixing on the RGB.}
\label{n1846_nuc_1}
\end{center}
\end{figure}

Stellar evolution calculations by \citet{lebzelter08} show that by assuming extra-mixing on the RGB along with an 
initial [O/Fe] = 0.20 dex, the predicted carbon and oxygen abundances match the 
observed M-star compositions. This result agrees well with our conclusions. Similar results were 
also presented by \citet{karakas10b} for the cluster's M-stars. In order to match the 
carbon and oxygen abundances in the C-stars, \citet{lebzelter08} estimated the effects of 
extra-mixing on the RGB as well as a moderate extra-mixing
in the late part of the AGB when the stars become C-rich. The theoretical models match both the 
observed M and C-star compositions as a result of the extra \iso{13}C brought to the surface 
by extra-mixing on the AGB. However it is unclear as to why extra-mixing starts on the AGB only when 
C/O $>$ 1. Nevertheless we also conclude that extra-mixing may be required on the AGB to explain the 
compositions of the C-stars. It is possible that this extra-mixing on the AGB is 
weak hot-bottom burning where the base of the convective envelope extends into the 
upper regions of the H-shell and some CN-cycling occurs. Since hot-bottom burning increases 
with stellar mass, and the AGB stars in NGC\,1846 are more massive than the 
AGB stars in NGC\,1978, this could explain why this effect is seen in NGC\,1846 and not 
NGC\,1978.

In Figure~\ref{n1846_nuc_1} we show the results for the $\log$ C/O versus
$\log$ \iso{12}C/\iso{13}C ratios for several additional model sequences. 
The scaled-solar model (black solid-line) 
does not match the C/O and the \iso{12}C/\iso{13}C ratios of the M-stars with the lowest C/O ratios. 
These stars show no C-enrichment and probably have a composition similar to (or that of) 
stars at the tip of the RGB. We find that the scaled-solar model matches the C/O 
and the \iso{12}C/\iso{13}C ratios of the other M-stars that show an enrichment 
of carbon caused by TDU, although their C/O $<$ 1. However, we find that this model does not reproduce the 
\iso{12}C/\iso{13}C of the C-stars. For the carbon-depleted model without extra-mixing 
(red dashed-line) we find that though the predicted C/O ratios lie in the range of the 
observed C/O ratios for the M and the C-stars, we cannot match the \iso{12}C/\iso{13}C ratio of 
the M-stars, where the post-FDU \iso{12}C/\iso{13}C ratio for this model is 22. Furthermore, 
the \iso{12}C/\iso{13}C ratio predicted for the C-stars is very far from the 
observed value. The carbon-depleted 
model with extra-mixing fits the observed C/O and the \iso{12}C/\iso{13}C ratios of the M-stars in the 
cluster (with a predicted post-FDU \iso{12}C/\iso{13}C = 12, refer Table~\ref{table3}). However, 
as for the previous models, this model also fails 
to reproduce the observational \iso{12}C/\iso{13}C ratio of the cluster's C-stars.

\subsubsection{Fluorine Abundances in NGC\,1846}

\begin{figure}
\begin{center}
\includegraphics[width=10cm,angle=270]{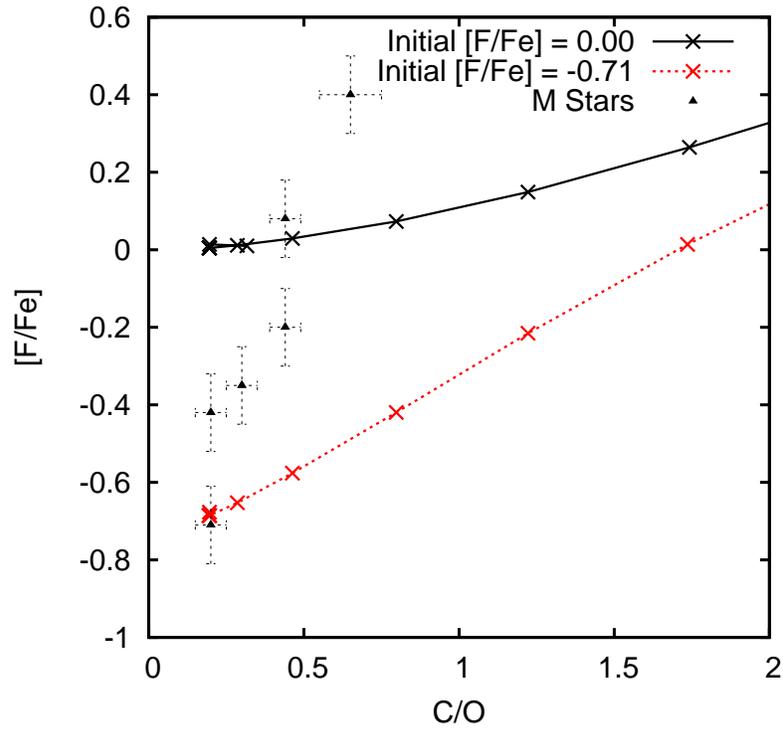} 
\caption{[F/Fe] versus C/O for NGC\,1846. The black solid 
line corresponds to the $\alpha$-enhanced model with a initial scaled-solar [F/Fe] = 0.0. 
The red dotted-line corresponds to the $\alpha$-enhanced model for which the 
initial [F/Fe] = $-$0.71. The black triangles denote the observed [F/Fe] 
values varying with C/O for the cluster's M-stars \citep{lebzelter08}.}
\label{n1846_F}
\end{center}
\end{figure}

An additional observational constraint in NGC\,1846 is the [F/Fe] ratio 
\citep{lebzelter08}. Observations from \citet{lebzelter08} show a steep increase in the estimated 
fluorine abundance with the C/O ratio. The black solid-line in Figure~\ref{n1846_F} shows the 
predicted [F/Fe] versus C/O ratios using our best fit model (the $\alpha$-enhanced 
model with extra-mixing) for the AGB stars in NGC\,1846. We also compute another model using the lowest 
observed [F/Fe] abundances \citep{lebzelter08} as a starting point, where the initial 
[F/Fe] = $-$0.71 (denoted using red dotted-lines in Figure~\ref{n1846_F}). 
We find that the increase in the predicted abundance of fluorine with the C/O ratio 
from both the theoretical models is shallower than the increase in the observed fluorine abundance 
for a given C/O ratio. Theoretical models by \citet{lebzelter08} also show a similar trend in
the [F/Fe] versus C/O ratios.

Flourine production in AGB stars is quite complicated: F is synthesized during 
thermal pulses via a combination of neutron, proton, and alpha capture reactions.
The most likely path for the production of fluorine is via \iso{14}N($\alpha,\gamma$)
\iso{18}F($\beta^{+}$)\iso{18}O(p,$\alpha$)\iso{15}N($\alpha,\gamma$)\iso{19}F reactions 
\citep{forestini92,mowlavi96}. The required \iso{15}N can be synthesized by the 
\iso{18}O(p,$\alpha$)\iso{15}N reaction owing to the presence of \iso{18}O and protons 
in the He intershell. \iso{18}O is produced by the $\alpha$-capture on 
\iso{14}N where \iso{14}N is found in the ashes of CNO cycling from the 
preceding H-burning stage. Further, the \iso{14}N(n,p)\iso{14}C reaction 
has a high cross section and can produce \iso{14}C and free protons. 
\iso{19}F production is enhanced by the inclusion of a \iso{13}C pocket via the release 
of free neutrons which come from the \iso{13}C($\alpha$,n)\iso{16}O reaction. At the end of each TDU episode, 
the convective envelope penetrates into the stable radiative intershell zone and 
a \iso{13}C-rich region can form in the top layers of the He-intershell 
as a result of the partial mixing of protons. The neutrons from the  
\iso{13}C($\alpha$,n)\iso{16}O reaction not only produce the s-process elements but also 
get captured by species such as \iso{14}N and \iso{26}Al; both of which are strong neutron absorbers 
and are produced by the H shell. Neutron captures on \iso{14}N then help 
increase the \iso{19}F production during a thermal pulse \citep[see, e.g.,][]{lugaro04}.

\begin{figure}
\begin{center}
\includegraphics[width=14cm,angle=0]{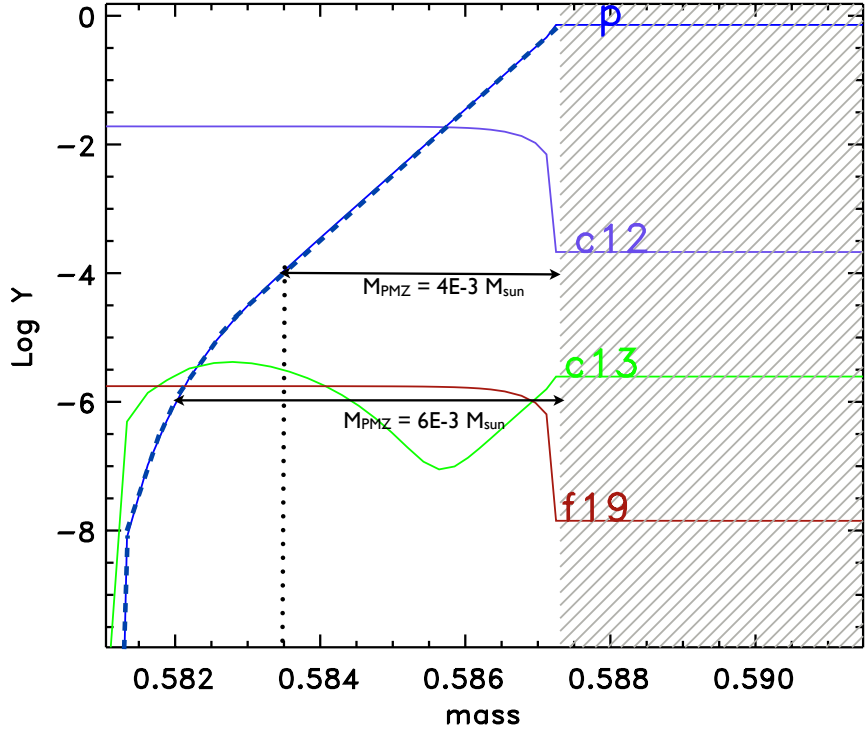} 
\caption {Proton profiles for the NGC\,1846 $\alpha$-enhanced model 
with the inclusion of partial mixing zones after the 4$^{\rm th}$ thermal pulse. The blue dashed-line shows the 
proton profile at the base of the convective envelope (shaded region) for 
M$_{\rm PMZ}$ = 6$\times$10$^{-3}$\Msun, where X$_{p} =$ 1$\times$10$^{-6}$. 
The black dotted-line shows schematically the lower edge of the partial 
mixing zone in a case where M$_{\rm PMZ}$ = 4$\times$10$^{-3}$\Msun, where 
X$_{p} =$ 1$\times$10$^{-4}$. Also shown are the composition 
profiles for \iso{12}C, \iso{13}C, and \iso{19}F. Note: The increse in \iso{13}C 
at $\approx$ 0.586 is due to the start of the formation of the \iso{13}C pocket.}
\label{pmz}
\end{center}
\end{figure}

In our best-fit model for the AGB stars in NGC\,1846, we artificially 
include a partial mixing zone at deepest extent of each TDU episode in the 
nucleosynthesis code using the same procedure as that explained in \citet{lugaro04} 
and \citet{karakas10a}, which is based on the method used by \citet{goriely00}. 
We insert the partial mixing zone at the deepest extent of the TDU because this is when 
a sharp discontinuity is produced between the convective envelope and 
the radiative intershell, which is a favourable 
condition for the occurrence of mixing. The partial mixing zone can be defined in terms of the 
proton profile. It is defined as the region where the abundance of protons drops exponentially 
from the envelope value to a fixed lower value (X$_{p}$). We experiment 
with two values of X$_{p} =$ 1$\times$10$^{-4}$ and X$_{p} =$ 1$\times$10$^{-6}$. For the 
size of the partial mixing zone we consider three values of M$_{\rm PMZ}=4\times$10$^{-3}$\Msun\, 
(i.e., $\sim$1/6 of the mass of the intershell at the deepest extent of TDU) and 
M$_{\rm PMZ} =$ 6$\times$10$^{-3}$\Msun\, (i.e., $\sim$1/4 of the mass 
of the intershell at the deepest extent of TDU), and M$_{\rm PMZ}=1.2\times$10$^{-2}$\Msun\, 
(i.e., most of the mass of the intershell during the last few TP's). Figure~\ref{pmz} 
shows M$_{\rm PMZ}$, X$_{p}$, and the proton profile in two cases. Note that M$_{\rm PMZ}$ 
stays constant with evolution along the AGB.

\begin{figure}
\begin{center}
\includegraphics[width=12cm,angle=270]{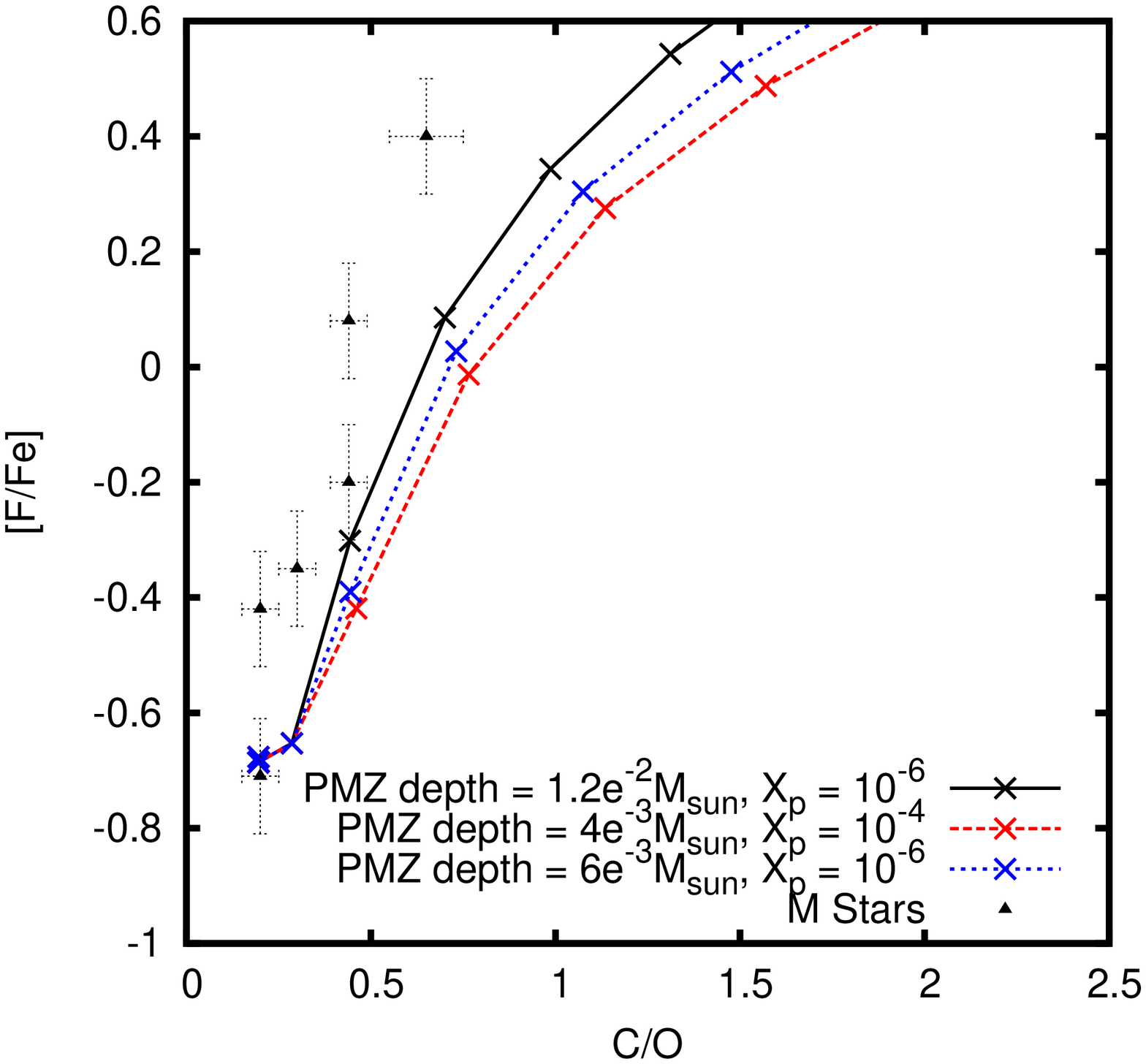} 
\caption{The effect of partial mixing zones (pmz) and proton-profiles (X$_{p}$) 
on the [F/Fe] versus C/O for NGC\,1846 $\alpha$-enhanced model. The black solid-line 
represents the model with M$_{\rm PMZ}$ = 1.2$\times$10$^{-2}$\Msun\, and 
X$_{p} =$ 1$\times$10$^{-6}$. The 
red dashed-line represents the model with M$_{\rm PMZ}$ = 4$\times$10$^{-3}$\Msun\, and 
X$_{p} =$ 1$\times$10$^{-4}$. The blue 
dotted-line represents the model with M$_{\rm PMZ}$ = 6$\times$10$^{-3}$\Msun\, and 
X$_{p} =$ 1$\times$10$^{-6}$. The black filled triangles indicate 
the observed F abundances of the M-stars \citep{lebzelter08}.}
\label{n1846_F2}
\end{center}
\end{figure}

Figure~\ref{n1846_F2} shows the predicted [F/Fe] versus C/O ratios for our best 
fit model for NGC\,1846 after the inclusion of partial mixing zones starting with an initial 
[F/Fe] $\sim$ $-$0.71. The red dashed-line represents the model where the partial 
mixing zone has a depth of M$_{\rm PMZ}$ = 4$\times$10$^{-3}$\Msun\, 
and a proton abundance limit, X$_{p} =$ 1$\times$10$^{-4}$. This model does not 
reproduce the observations. We experiment by extending the depth of the partial mixing zone 
to M$_{\rm PMZ}$ = 6$\times$10$^{-3}$\Msun\, and X$_{p} =$ 1$\times$10$^{-6}$ denoted by the 
blue dotted-line in Figure~\ref{n1846_F2}. The model shows that a little more F is produced 
at a given C/O ratio than the 
previous case, however, the  fluorine abundance does not match the observations. 
The black solid-line in Figure~\ref{n1846_F2} 
represents the best-fit model of NGC\,1846 with a M$_{\rm PMZ}$ = 1.2$\times$10$^{-2}$\Msun\, and 
X$_{p} =$ 1$\times$10$^{-6}$. This model reproduces the observed F abundances fairly well. 
This implies that a M$_{\rm PMZ}$ = 1.2$\times$10$^{-2}$\Msun\, or higher would be required to 
reproduce the observations: note that the total model intershell mass is reduced to 
$\sim$1.5$\times$10$^{-2}$\Msun\, at the tip of the AGB. However, 
this situation is rather speculative as $s$-process 
models require smaller $^{13}$C pockets in order to match observations \citep{gallino98,axel07}.

We conclude that our models can reproduce the observations but only with partial mixing zones 
that are likely to be larger than those required for $s$-process studies, making this a speculative result. 
We also note that the observed abundances were estimated based on the single blended HF 
line for the M-stars which also adds to the uncertainity in the observed F abundance.

\subsection{Evolution and Nucleosynthesis model results for NGC\,419}
\label{sec:419results}

\begin{figure}
\begin{center}
\includegraphics[width=15cm,angle=0]{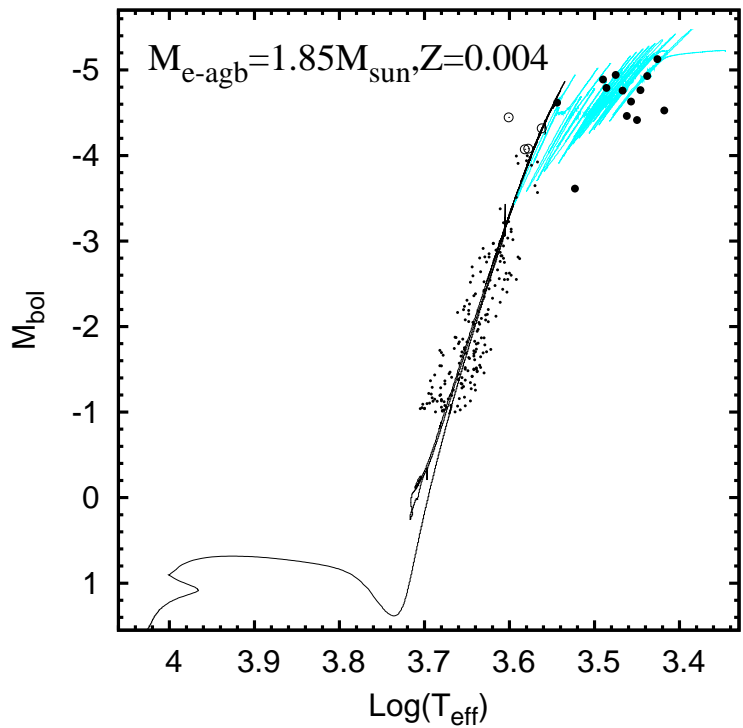} 
\caption{Evolutionary track for NGC\,419's scaled-solar model. The cyan/grey lines 
indicates that the star is C-rich. The black 
open circles represent the observed positions of the 
M-stars in NGC\,419, the black filled circles represent the 
observed positions of the C-stars in NGC\,419, and the small black 
dots represent the non-variable red-giants in the cluster \citep{kamath10}. 
Note that the low luminosity of the C-star indicates that 
the star is currently in a post-flash luminosity dip.
}
\label{n419}
\end{center}
\end{figure}

The last cluster in our study is the SMC cluster NGC\,419. 
In Figure~\ref{n419} we show the theoretical evolutionary track of the scaled-solar model 
and the positions of the AGB variables and the non-variable red-giants in NGC\,419 \citep{kamath10}. 
The AGB evolutionary track is clearly in good agreement with the observed star positions. 
Here we only consider a scaled solar initial composition owing to the 
lack of information on the abundances for stars 
in this cluster. We use a ZAMS model with an initial mass of 1.91\Msun, $Z = 0.004$, and $Y = 0.25$. 
This results in an E-AGB mass of 1.85\Msun\, which agrees well with the estimated 
pulsation mass for the AGB variables in this cluster \citep{kamath10}. 
We find that an $\alpha$ of 1.74 is required to reproduce the 
giant branch temperatures: this is similar to that for NGC\,1846 which is of a similar 
initial mass (Table~\ref{table1}). An overshoot of $N_{\rm ov}$ = 2.10 is required on the 
AGB to reproduce the observed M/C transition luminosity ($M_{\rm bol} = -4.5$). To match the observed AGB-tip 
luminosity, the super-wind mass-loss rate needs to begin at a pulsation period of 790 days. 
Owing to a lower metallicity and a higher mass, the AGB stars experience more TPs (19) compared to
the two other cluster AGB stars ($\sim$ 15 each). The greater number of TPs 
leads to a large amount of He-shell burning material being mixed
into the envelope and this is demonstrated by the high final C/O and 
\iso{12}C/\iso{13}C ratios (Table~\ref{table3}). The envelope mass when convergence failed 
is $\sim$ 0.29\Msun which implies that 
the models star is unlikely to experience any further TPs or TDU episodes since the mass-loss rate at 
this stage is 9.40$\times$10$^{-6}$\Msun/year, and the interpulse period is $\sim$ 1$\times$10$^{5}$ years.

\subsection{The effect of C/O ratio on $T_{\rm eff}$ of AGB stars}

\begin{figure}
\begin{center}
\includegraphics[width=14cm,angle=0]{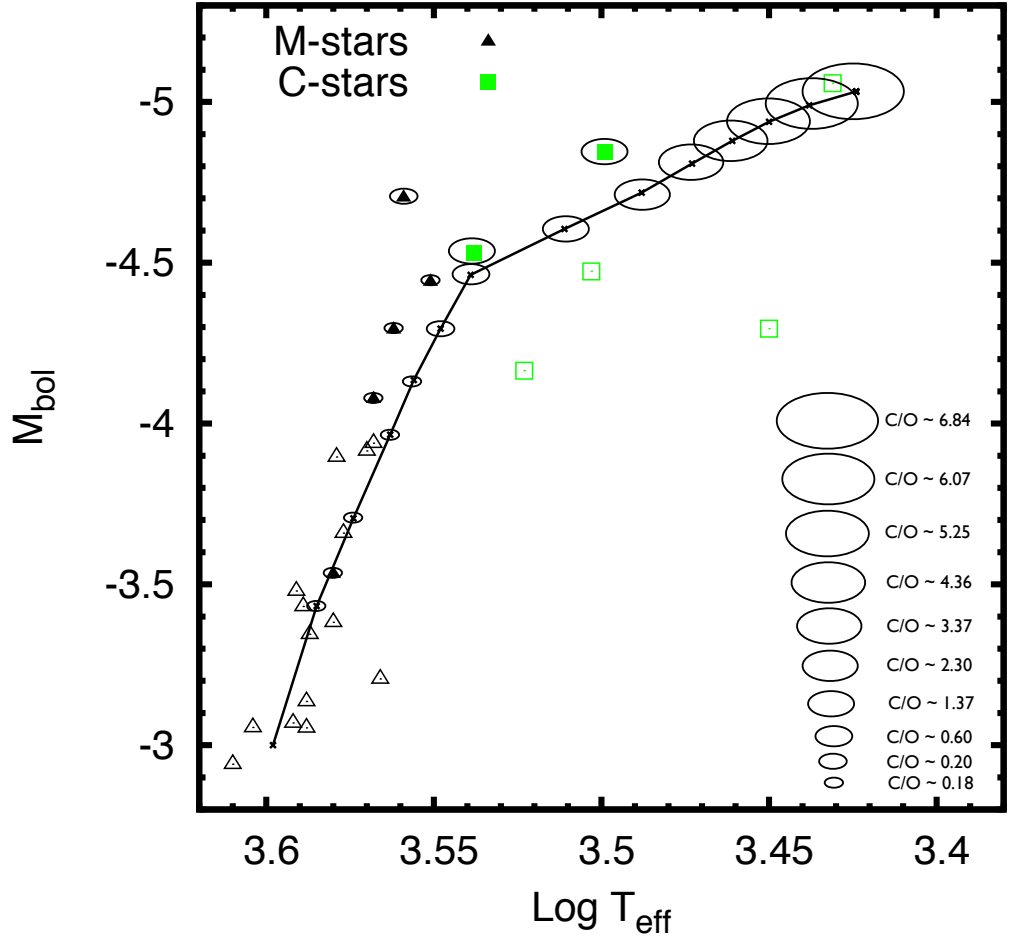} 
\caption{$M_{\rm bol}$ versus $\log$ $T_{\rm eff}$ for NGC\,1978. The black solid-line 
corresponds to the best-fit model. The ellipses mark the predicted C/O ratios at the end of each TDU 
episode. A grid of the C/O estimates is shown in the bottom-right corner. The filled black triangles 
indicate the M-stars for which observational C and O abundances exist. The filled green 
squares indicate the C-stars for which observational C and O abundances exist. The size of the 
ellipse around the filled symbols indicates the C/O ratio of the star. The open black 
triangles and green squares indicate the other M-stars and C-stars, respectively, in the cluster.}
\label{1978test}
\end{center}
\end{figure}

\begin{figure}
\begin{center}
\includegraphics[width=14cm,angle=0]{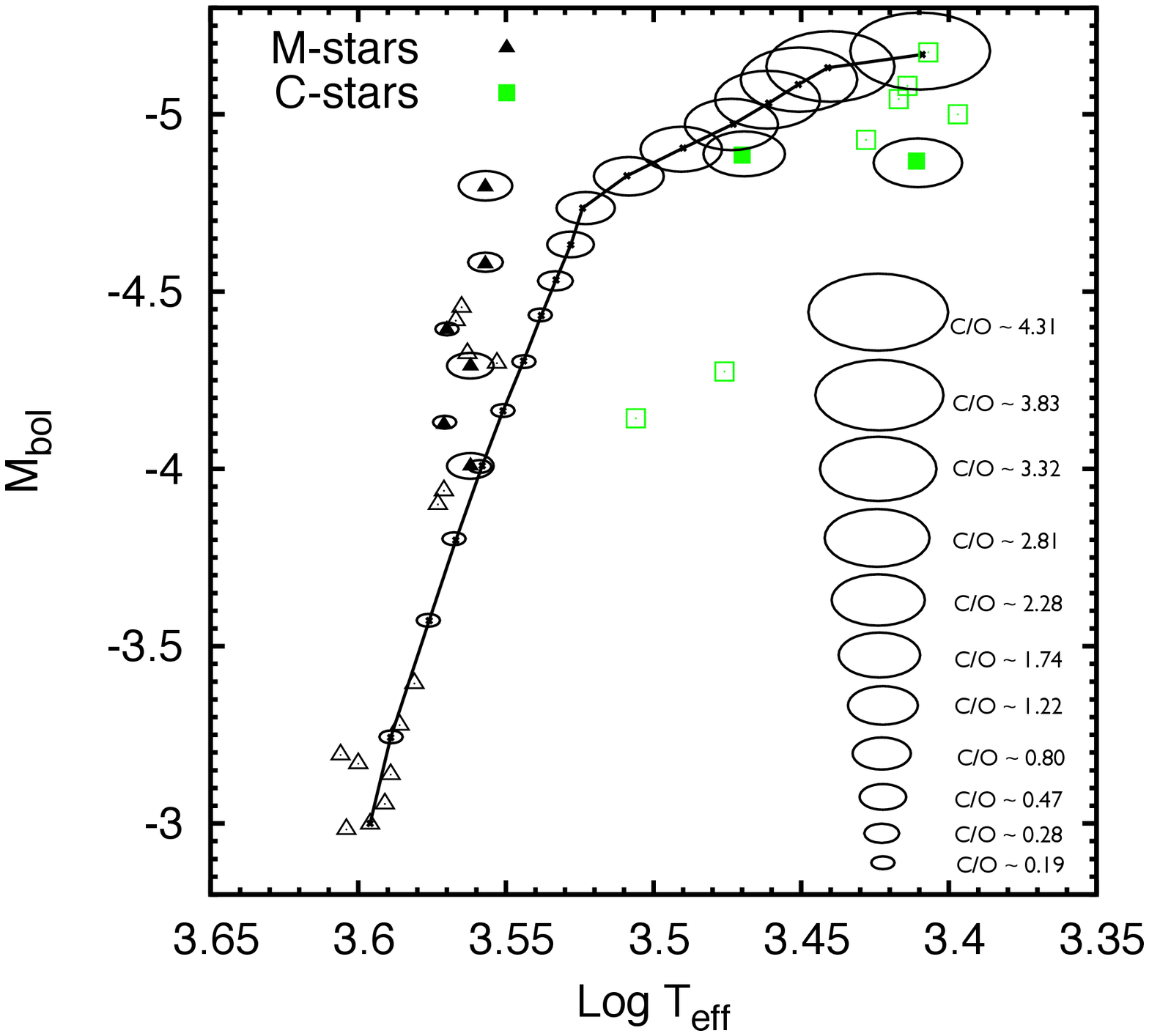} 
\caption{The same as Figure~\ref{1978test} for NGC\,1846's best-fit model.} 
\label{1846test}
\end{center}
\end{figure}

Figure~\ref{1978test} and Figure~\ref{1846test} show the variation of $M_{\rm bol}$ with 
$\log$ $T_{\rm eff}$ measured at the interpulse luminosity maximum for each of the TP's 
for the two LMC clusters NGC\,1978 and NGC\,1846 (black solid-lines). The size of the ellipses
on the solid lines denotes the C/O ratio after each TDU episode, 
as indicated in the grid on the bottom-right corner of the figures. Also shown are the AGB variables that have 
observational C/O estimates (Lederer et al. 2009 for NGC\,1978, and Lebzelter et al. 2008 for 
NGC\,1846), with the M-stars depicted as black filled triangles and C-stars as green 
filled squares. The size of the ellipse around each of the filled symbols 
indicates the observational C/O ratio of the star. The AGB variables without any abundance 
information are denoted as open symbols (M-stars 
as black open triangles and C-stars as green open squares). For most of the observational data, the 
$M_{\rm bol}$ and the $\log$ $T_{\rm eff}$ is taken from \citet{kamath10} for NGC\,1978 and 
\citet{lebzelter07} for NGC\,1846 and for those stars not in these studies, the 
$M_{\rm bol}$ and the $\log$ $T_{\rm eff}$ is taken from \citet{lederer09b} for NGC\,1978 and 
\citet{lebzelter08} for NGC\,1846.

Figures~\ref{1978test} and \ref{1846test} shows that 
for the M stars with luminosities $-4 < M_{\rm bol} < -3$, $T_{\rm eff}$ of the models
matches the observed values of $T_{\rm eff}$ well.  This is because the mixing
length was adjusted to provide a match between model and observed $T_{\rm eff}$ values
at these luminosities.  At more luminous $M_{\rm bol}$, 
the model $T_{\rm eff}$ is slightly cooler than the observations. This can be attributed to the 
effect of the constant mixing length that we use in our models. Previous studies 
\citep[e.g.][]{lebzelter07} have shown that at high luminosities the mixing 
length needs to be increased slowly with luminosity 
in order to reproduce the correct slope for the theoretical giant branch.

Once the M/C transition luminosity is reached, it can be seen that the
AGB slope of the models suddenly decreases and the stars become increasingly cooler
with increasing C/O ratio.  The few C-stars with observed C/O ratios generally
show the same movement to lower $T_{\rm eff}$ values as seen in the models, 
although the correspondence
between $T_{\rm eff}$ and C/O ratio is not clear.  The larger sample of C stars
that do not have C/O measurements also lie cooler than the sequence
of O-rich AGB stars or its extrapolation to higher luminosities.  However,
these stars do not necessarily lie directly on the sequences shown in 
Figures~\ref{1978test} and \ref{1846test} for models
at the interpulse luminosity maximum because of the variation of luminosity
over the thermal pulse cycle.  This is best seen in Figures~\ref{n1978}, \ref{n1846_HR}
and \ref{n419}.  In summary, the increase in C/O ratio above unity causes
$T_{\rm eff}$ to decrease but it is hard to determine
from the current observational data whether the magnitude of this effect
is reproduced accurately by the models.

\section{Discussion and Conclusions} 
\label{sec:DnC}

In this paper we have presented new stellar models for AGB stars in 
NGC\,1978, NGC\,1846, and NGC\,419. These are three clusters whose well-defined observational characteristics 
provide strong constraints on evolution models. 
The stellar evolution  models are constrained to reflect the observed AGB 
pulsation mass, cluster metallicity, giant branch effective temperature, 
M/C transition luminosity and the AGB-tip luminosity. 

Stellar models are affected by major uncertainties including mass loss, 
convection (e.g., mixing length), depth and efficiency of TDU, and 
non-convective mixing. In our study we 
try to constrain these uncertainties. In order to produce the correct 
AGB-tip luminosities, the \citet{vw93} mass-loss 
prescription had to be modified so that the super-wind mass-loss 
rate commences at a period of $\sim$ 710 $-$ 790 days as opposed to the 500 days period employed 
by \citet{vw93}. The mixing length parameter ($\alpha$) required to fit the observed giant branch 
temperature for each of the three clusters shows values similar to the $\alpha$'s found 
when calculating a standard solar model \citep[e.g.,][although our
values are slightly lower than their $\alpha \approx 2.1$]{piersanti07}.

Overshoot at the base of the convective envelope of $\sim$ 1 $-$ 3 pressure 
scale heights was required to obtain the observed 
luminosity of the M/C transition. This results  
in higher final C/O ratios of about $\sim$ 7.5 $-$ 8.0 for the LMC cluster NGC\,1978, 
$\sim$ 3.7 $-$ 4.8 for the LMC cluster NGC\,1846, and $\sim$ 11.9 for the SMC cluster NGC\,419. 
The C/O ratios of planetary nebulae (PNe) put upper limits on the C/O ratios for the LMC and SMC. 
In the LMC, \citet{stanghellini05} found that a C/O of 2 is typical for  non-bipolar PN and in the SMC 
a C/O = 4 is typical for non-bipolar PN \citep{stanghellini09}. This indicates that the models 
show a higher level of C-enrichment when compared to the observations. It is worth mentioning 
that the model with an enhanced C and O intershell composition, computed to reproduce the 
observational C and O abundances of the M and C-stars in the LMC cluster NGC\,1978 
(refer Section~\ref{sec:1978results}), has a final surface C/O ratio of C/O $\approx$ 3.00, 
in good agreement with the LMC PN data.

\begin{figure}
\begin{center}
\includegraphics[width=14cm,angle=0]{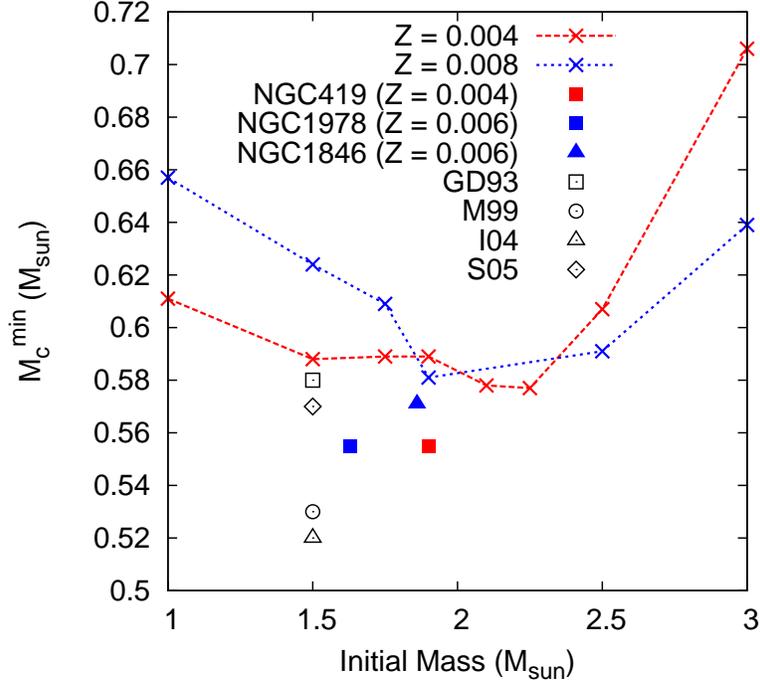} 
\caption{The $M_{c}^{\rm min}$ plotted against initial mass for the Z = 0.008 
(blue dotted-line and points) and Z = 0.004 (red dashed-line and points) models 
from \citep{karakas02}. Also plotted are the average $M_{c}^{\rm min}$ values for the 
models in each of the three clusters. The values of $M_{c}^{\rm min}$ derived for 1.5\Msun\, 
models in synthetic AGB calculations designed to reproduce the LMC C-star 
luminosity functions are shown for \citet{dejong93} (GD93, open square), 
\citet{marigo99} (M99, open circle), \citet{izzard04b} (I04, open triangle), and 
\citet{stancliffe05a} (S05, open diamond).}
\label{Mcmin_Mi}
\end{center}
\end{figure}

\begin{figure}
\begin{center}
\includegraphics[width=14cm,angle=0]{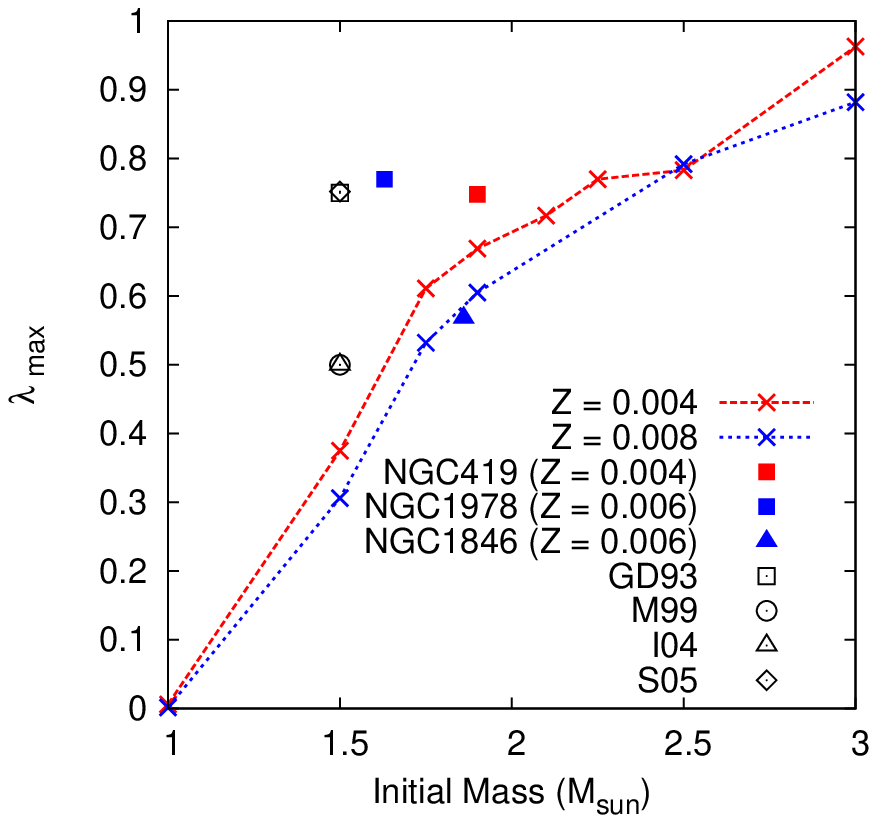}
\caption{Same as Fig.~\ref{Mcmin_Mi} but showing the variation of  $\lambda_{\rm max}$ 
with initial mass for full evolution calculations and the values for the synthetic 
calculations. Note that the $\lambda$ values from GD93 and M99 are constant values, while 
$\lambda$ values vary in I04 and S05 and the value plotted is the maximum value.}
\label{Lambda_Mi}
\end{center}
\end{figure}

Another consequence of the overshoot employed in our models is that the 
average $\lambda$ values from our full AGB evolution calculations lie in the range 0.40 to 0.70 for our 
best fitting models. The first dredge-up episodes occur for core masses of 
$M_{c}^{\rm min}$ $\approx 0.56 - 0.58\Msun$.
Our values can be compared with evolution calculations that do not use 
any overshoot at the base of the convective envelope. For direct 
comparison we use the models of \citet{karakas02}. 
Figure~\ref{Mcmin_Mi} and Figure~\ref{Lambda_Mi} compares our
predicted $M_{c}^{\rm min}$ and $\lambda_{\rm max}$ (from 
Table 2) as a function of the initial mass and metallicty to
the $Z=0.008$ and $Z=0.004$ model values from \citet{karakas02}. We
find that, as expected, our models find TDU at smaller core masses
than the models of \citet{karakas02} without overshoot. Similarly,
our $\lambda_{\rm max}$ values are higher.

The C-star luminosity function (CSLFs) in the MCs have been used to calibrate
the onset and efficiency of the TDU in synthetic AGB evolution
calculations \citep{dejong93,marigo99,izzard04b,stancliffe05a}. The values derived
in these four studies are also shown in Figure~\ref{Mcmin_Mi} and
Figure~\ref{Lambda_Mi}.  The $M_{c}^{\rm min}$ values span a wide
range downward from the values found by \citet{karakas02} and 
they bracket our values. The \citet{izzard04b} and \citet{marigo99} 
value seem too small to be consistent with our results. In the synthetic calculations 
by \citet{dejong93} and \citet{marigo99}, $\lambda$ 
is a constant parameter that is altered to fit the CSLFs but $\lambda$ varies 
in the models of \citet{izzard04b} and \citet{stancliffe05a}. Given the range of 
masses and metallicities involved in the models and observations, it is 
difficult to use our derived $\lambda$ values to make precise comments on the 
validity of the $\lambda$ values usined in the synthetic models. 
Observations from clusters with a wider range of
initial masses would be useful for constraining these parameters 
as a function of mass and metallicity.

In Table~\ref{table2} we give the final core and total masses for the
stellar models. Given that we expect no further TPs, the final core 
masses are approximately the final masses, and these can be compared 
to white dwarf masses. Our predicted final masses are $\approx 0.59-0.64\Msun$,
right at the peak of the distribution of white dwarf masses in the
Milky Way Galaxy \citep{ferrario05,liebert05}. Even though these white
dwarf masses are derived from a different stellar population (notably
one more metal rich), there is a strong consistency between our predicted
masses and the peak of the observed range.

The main focus of our study is to establish a theoretical explanation for the observed C, O, 
and F abundances (F abundances only for NGC\,1846) in the three clusters. 
\citet{lederer09b} had difficulties finding a theoretical explanation for the 
NGC\,1978 M and C-type AGB stars. Our scenario uses a 
carbon-depleted model with extra-mixing on the RGB along with an intershell 
enhanced in both \iso{12}C and \iso{16}C to explain the composition of the 
stars in NGC\,1978.  This scenario is somewhat speculative because it
is not clear if AGB stars experience convective overshoot between the
flash-driven convective region and C-O core, even though this mixing
likely occurs in post-AGB stars to produce the PG 1159 stars. 
For NGC\,1846 we find that an $\alpha$-enhanced 
initial composition coupled with extra-mixing on the RGB followed by standard thermally pulsing 
AGB evolution reproduces the observed abundance 
patters of the cluster M-stars. We find that extra-mixing , perhaps the beginning of 
hot-bottom burning, is also required on the AGB 
in order to fit the observational data for the C-stars though we do not model
this in our study.  This conclusion was also reached by \citet{lebzelter08}. For the M-stars 
in NGC\,1846 we insert partial mixing zones which turn into the \iso{13}C pocket 
in our post-processing nucleosynthesis models in order to try and re-produce 
the observed F abundances. We find that a partial mixing zone
that is a similar size (in mass) to the intershell region at the
end of the TP-AGB will reproduce the observed F abundance variation with C/O. 
This result is in contradiction to $s$-process studies
that require smaller $^{13}$C pocket masses to reproduce
the observed $s$-process abundance distribution \citep{gallino98,axel07}.

For NGC\,419 we have predicted C and O abundances based on a scaled-solar model and these 
results can be compared to the observational abundances once they have been derived.

\acknowledgments

DK thanks Dougal Mackey and John Norris for helpful comments on the LMC clusters. 
AIK and DK thank Carlos Iglesias and Forrest Rogers for help with
obtaining  OPAL tables, and Richard Stancliffe for discussions on mixing in 
giant stars. This work benefited from the support of the 
NCI National Facility at the ANU.

\clearpage



\begin{table}
 \begin{center}
  \caption{Parameters used to construct the stellar models for 
the three target clusters -- NGC\,1978, NGC\,1846, and NGC\,419.
  \label{table1}}
  \vspace{1mm}
  \newcommand{\alfe}{[$\alpha$/Fe]}
  \newcommand{\cfe}{[C/Fe]}
   \begin{tabular}{lccccccccc}
   \tableline\tableline
Model Type & $M_{ZAMS}$ & Z & $\log$ $T_{\rm eff,4}$ & $\alpha$ & 
$M_{\rm e-agb}$ & $N_{\rm ov}$ & $M_{\rm bol}^{\rm
M/C}$ & $P$ & $M_{\rm bol}^{\rm AGB-tip}$ \\
           &  (\Msun)     & &   &          &
   (\Msun)        & (H$_{p}$)                        &                
                       & (days) & \\
\tableline
\multicolumn{10}{c}{NGC\,1978} \\ \tableline
Scaled-Solar      & 1.63 & 0.006 & 3.564 & 1.90 & 1.56 & 2.54 & $-$4.51 & 790 & $-$5.08\\
C-Depleted        & 1.63 & 0.006 & 3.563 & 1.90 & 1.56 & 2.75 & $-$4.47 & 790 & $-$5.06\\
$\alpha$-Enhanced & 1.63 & 0.006 & 3.564 & 1.90 & 1.56 & 3.00 & $-$4.50 & 790 & $-$5.05\\
\tableline
\multicolumn{10}{c}{NGC\,1846} \\ \tableline
Scaled-Solar      & 1.86 & 0.006 & 3.560 & 1.74 & 1.80 & 1.05 & $-$4.73 & 710 & $-$5.18\\
C-Depleted        & 1.86 & 0.006 & 3.558 & 1.74 & 1.80 & 1.05 & $-$4.77 & 710 & $-$5.18\\
$\alpha$-Enhanced & 1.86 & 0.006 & 3.559 & 1.74 & 1.80 & 1.41 & $-$4.78 & 710 & $-$5.17\\
\tableline
\multicolumn{10}{c}{NGC\,419} \\ \tableline
Scaled-Solar      & 1.90 & 0.004 & 3.574 & 1.74 & 1.85 & 2.10 & $-$4.49 & 790 & $-$5.23\\
\tableline \tableline
\tablenotetext{}{Note: $\log$ $T_{\rm eff,4}$ is the computed $\log$ $T_{\rm eff}$ 
for the M-stars on the AGB at $M_{\rm bol} = -4.00$, 
$N_{\rm ov}$ denotes the overshoot parameter in pressure scale heights, 
H$_{p}$ (see Section~\ref{sec:Non-std-intershell}). 
$M_{\rm bol}^{\rm M/C}$ denotes the computed bolometric luminosity at the M/C transition. 
$P$ is the pulsation period where the superwind mass-loss rate is reached. 
$M_{\rm bol}^{\rm AGB-tip}$ denotes the computed AGB-tip bolometric luminosity.}
 \end{tabular} 
 \end{center}
\end{table}

\clearpage

\begin{table}
 \begin{center}
  \caption{Characteristic parameters of the AGB models.
 \label{table2}}
  \vspace{1mm}
  \newcommand{\alfe}{[$\alpha$/Fe]}
  \newcommand{\cfe}{[C/Fe]}
   \begin{tabular}{lccccccccc}
   \tableline\tableline
Model Type & TPs &  $M_{c}^{\rm min}$ & $\lambda_{\rm max}$  & $\lambda_{\rm avg}$ & $M_{\rm dredge}$ & 
$T_{\rm Heshell}^{\rm max}$ & $M_{\rm c}$(f) &
$M{\rm tot}$(f) & $\tau$$_{\rm ip}$(f) \\ 
           & &  (\Msun)     &   &          & (\Msun) & (10$^{6}$K) &
   (\Msun)        & (\Msun) & (10$^{5}$years)\\

\tableline
\multicolumn{10}{c}{NGC\,1978 - $M_{ZAMS}$ = 1.63, Z = 0.006}\\ \tableline
Scaled-Solar      & 15 & 0.558 & 0.720 & 0.588 & 0.079 & 273 & 0.605 & 0.655 & 1.209\\
C-Depleted        & 13 & 0.555 & 0.773 & 0.653 & 0.085 & 272 & 0.599 & 0.856 & 1.398\\
$\alpha-$Enhanced & 13 & 0.552 & 0.817 & 0.655 & 0.097 & 272 & 0.594 & 0.647 & 1.392\\
\tableline
\multicolumn{10}{c}{NGC\,1846 - $M_{ZAMS}$ = 1.86, Z = 0.006} \\ \tableline
Scaled-Solar      & 16 & 0.569 & 0.551 & 0.429 & 0.051 & 279 & 0.635 & 0.933 & 1.021\\
C-Depleted        & 18 & 0.573 & 0.545 & 0.433 & 0.051 & 280 & 0.636 & 0.749 & 0.978\\
$\alpha-$Enhanced & 17 & 0.571 & 0.609 & 0.487 & 0.060 & 279 & 0.630 & 0.705 & 1.056\\
\tableline
\multicolumn{10}{c}{NGC\,419 - $M_{ZAMS}$ = 1.90, Z = 0.004} \\ \tableline
Scaled-Solar      & 19 & 0.555 & 0.748 & 0.608 & 0.118 & 279 & 0.625 & 0.914 & 0.997\\
\tableline \tableline
  \end{tabular} 
 \end{center}
\end{table}

\clearpage

\begin{table}
 \begin{center}
  \caption{Initial, post-FDU, and final C/O and \iso{12}C/\iso{13}C ratios 
for the three clusters. AGB-tip [F/Fe] abundances from 
the stellar models are also listed. All abundances are by number.
  \label{table3}}
  \vspace{1mm}
  \newcommand{\alfe}{[$\alpha$/Fe]}
  \newcommand{\cfe}{[C/Fe]}
  \newcommand{\coin}{C/O$_{\rm i}$}
  \newcommand{\copfdu}{C/O$_{\rm P-FDU}$}
  \newcommand{\cofin}{C/O$_{\rm f}$}
  \newcommand{\isoci}{\iso{12}C/\iso{13}C$_{\rm i}$}
  \newcommand{\isocpfdu}{\iso{12}C/\iso{13}C$_{\rm P-FDU}$}
  \newcommand{\isocf}{\iso{12}C/\iso{13}C$_{\rm f}$}
  \newcommand{\ffei}{[F/Fe]$_{\rm i}$}
  \newcommand{\ffef}{[F/Fe]$_{\rm f}$}
   \begin{tabular}{lcccccccc}
   \tableline\tableline
Model Type& EM\tablenotemark{a} & \coin\tablenotemark{b} & \copfdu\tablenotemark{b} & \cofin\tablenotemark{b} & \isoci & \isocpfdu &\isocf & \ffef \\ \tableline
\multicolumn{9}{c}{NGC\,1978} \\ \tableline
Scaled-Solar & No  & 0.501 & 0.333 & 8.258 & 89.4 & 23.4 &  691.2 & 1.147 \\
C-Depleted   & No  & 0.282 & 0.185 & 7.561 & 89.4 & 22.7 & 1106.2 & 1.078 \\
C-Depleted   & Yes & 0.282 & 0.184 & 7.601 & 89.4 & 12.8 &  654.0 & 1.080 \\
C-Depleted$^{\rm c}$   & No  & 0.281 & 0.184 & 7.566 & 50.0 & 19.3 & 948.4 & 1.078 \\
$\alpha-$Enhanced  & No  & 0.282 & 0.184 & 8.038 & 89.4 & 22.7 & 1175.3 & 1.369 \\
$\alpha-$Enhanced  & Yes & 0.282 & 0.186 & 8.079 & 89.4 & 12.8 &  684.4 & 1.371 \\
\tableline
\multicolumn{9}{c}{NGC\,1846} \\ \tableline
Scaled-Solar    & No  & 0.501 & 0.309 &  4.700 & 89.4 & 21.9 &  386.2 & 0.740 \\
C-Depleted & No  & 0.282 & 0.173 & 4.480 & 89.4 & 21.8 & 651.5 & 0.741 \\
C-Depleted & Yes & 0.282 & 0.167 & 4.489 & 89.4 & 11.8 &  378.6 & 0.743\\
$\alpha-$Enhanced  & No  & 0.316 & 0.196 &  4.821 & 89.4 & 21.8 & 621.1 & 0.996 \\
$\alpha-$Enhanced  & Yes & 0.316 & 0.195 &  4.832 & 89.4 & 13.8 & 400.8 & 0.998 \\
\tableline
\multicolumn{9}{c}{NGC\,419} \\ \tableline
Scaled-Solar    & No  & 0.501 & 0.299 & 11.954 & 89.4 & 21.6 & 1101.7 & 1.393 \\
\tableline \tableline
  \tablenotetext{a}{This indicates if extra-mixing is assumed to occur on the RGB.}
 \tablenotetext{b}{The notations 'i', 'P-FDU', and 'f' denote the initial, post-FDU, and final abundance values, respectively.}
\tablenotetext{c}{This represents the carbon-depleted model with an initial \iso{12}C/\iso{13}C = 50 (see Section~\ref{sec:1978results} for details).}
 
  \end{tabular} 
\end{center}
  \end{table}

\clearpage
  




\bibliographystyle{apj}

\bibliography{mnemonic,devlib}

\end{document}